\pdfoutput=1

\documentclass[sigconf]{acmart}
\renewcommand\footnotetextcopyrightpermission[1]{} 
\AtBeginDocument{%
  \providecommand\BibTeX{{%
    \normalfont B\kern-0.5em{\scshape i\kern-0.25em b}\kern-0.8em\TeX}}}

\usepackage{multirow}
\usepackage{lineno}
\usepackage{subcaption}
\usepackage{caption}
\usepackage{booktabs}
\usepackage{threeparttable}

\def\ie{\emph{i.e.}}
\def\eg{\emph{e.g.}}
\def\mosen{\emph{MoSen}}

\setcopyright{acmlicensed}

\begin{document}

\title[MoSen Activity Modelling]{MoSen: Activity Modelling in Multiple-Occupancy Smart Homes}

\pagestyle{plain}

\author{Yuting Zhan, Hamed Haddadi}
\affiliation{%
 \institution{Imperial College London}
 }

\begin{abstract}

Smart home solutions increasingly rely on a variety of sensors for behavioral analytics and activity recognition to provide context-aware applications and personalized care. 
Optimizing the sensor network is one of the most important approaches to ensure the classification accuracy and system's efficiency.
However, the trade-off between the cost and performance is often a challenge in real deployments, particularly for multiple-occupancy smart homes or care homes.


In this paper, using real indoor activity and mobility traces, floor plans, and synthetic multi-occupancy behavior models, we evaluate several multi-occupancy household scenarios with 2-5 residents.
We explore and quantify the trade-offs between the cost of sensor deployments and expected labeling accuracy in different scenarios.
Our evaluation across different scenarios show that the performance of the desired context-aware task is affected by different localization resolutions, the number of residents, the number of sensors, and varying sensor deployments.
To aid in accelerating the adoption of practical sensor-based activity recognition technology, we design \emph{MoSen}, a framework to simulate the interaction dynamics between sensor-based environments and multiple residents.
By evaluating the factors that affect the performance of the desired sensor network, we provide a sensor selection strategy and design metrics for sensor layout in real environments.
Using our selection strategy in a 5-person scenario case study, we demonstrate that \emph{MoSen} can significantly improve overall system performance without increasing the deployment costs.

\end{abstract}

\maketitle
\section{Introduction}
Human activity recognition (HAR) is the central task to many intelligent systems such as smart homes~\cite{bianchi2019iot}, long-term healthcare~\cite{subasi2018iot}, personal  robotics~\cite{rodriguez2020context}, assisted living~\cite{patel2019sensor}, and human-computer interaction~\cite{hassan2018robust}. 
Current works illustrate that human activity can be recognized using two main approaches, namely video-based~\cite{jalal2017robust} and sensor-based~\cite{cook2013transfer}.
Video-based activity recognition utilizes cameras to capture or record individuals' motions~\cite{jalal2017robust}, while sensor-based systems leverage the wearable or ambient sensors to understand the movements of the subjects and the interactions between people and the environment~\cite{tapia2004activity}.
While video-based approaches are often privacy-invasive, the sensor-based systems, which we focus on in this paper, are often more privacy-friendly and take advantage of their pervasiveness~\cite{wang2019deep}.
Currently, more and more sensors are getting embedded into our ambient environment, wearable electrical products, and intelligent appliances to aid with sensor-based activity recognition systems.
The multi-modal sensor data enables the system to receive rich context information and to have the capability to process personalized behavioral analytics and provide context-aware applications~\cite{bianchi2019iot}.


While multitudes of sensors extend the variety of information that could be received, 
the heterogeneity of the devices~\cite{stisen2015smart} and the increasing number of the residents~\cite{benmansour2015multioccupant, erickson2014occupancy} complicate the data collection system in real settings.
Even for \emph{single-occupancy scenarios}, where only a single individual in the single space, the diversity of sensor settings or floorplans could affect the overall performance of sensor networks. 
Importantly, sensor networks designed for single-occupancy houses are never deployed in identical settings, and sensor selection in each system is diverse, varying from commercial products to self-built devices~\cite{alemdar2013aras,cook2012casas,van2008accurate, cerpa2001habitat}. 
The price, stability, precision and coverage range of different sensors affect the implementation and performance of sensor-based systems~\cite{cerpa2001habitat}.
It is difficult to find a uniform sensor integration system flexible to distinct homes, especially when the homes might have more than one resident, referring to the \textit{multi-occupancy scenarios} in this paper.
Prior research has already specified the significance of multi-occupancy scenarios, but the complexity in the ongoing sensor networks and unknown uncertainties impede the real implementation of the sensor network and further analysis~\cite{adib2015multi, bocca2013multiple, benmansour2015multioccupant, mokhtari2018multi, wang2019deep}. 
Hence, when designing a specific sensor network to the target home, especially in multi-occupancy scenarios, people need to take the real floorplan, the number of residents, sensor density, and device's resolution into considerations.

\begin{figure*}[t]
     \centering
     \hfill
     \begin{subfigure}[c]{0.32\textwidth}
         \centering
         \includegraphics[width=\textwidth]{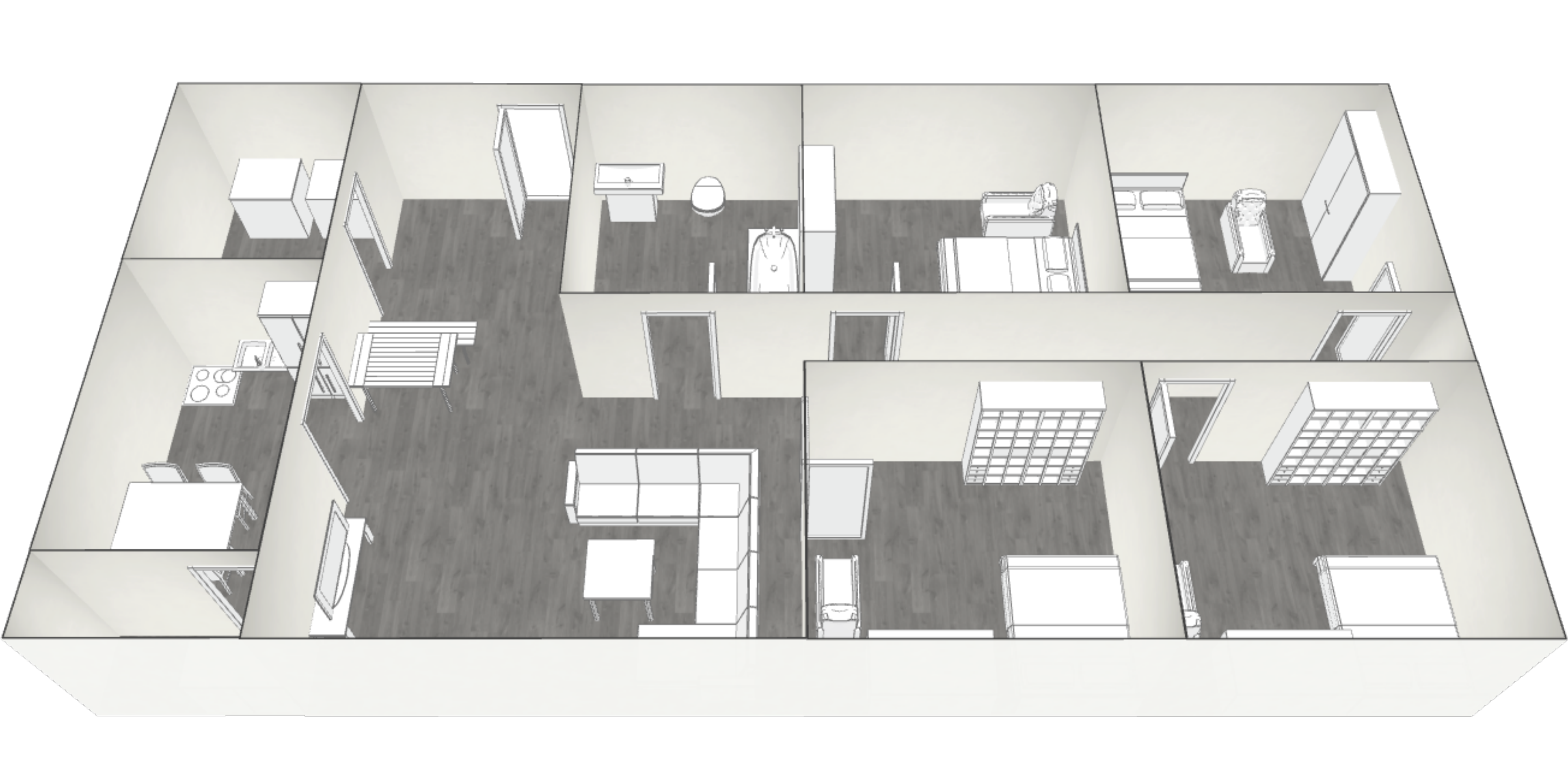}
         \caption{3D view}
         \label{fig:3D}
     \end{subfigure}
     \hfill
     \begin{subfigure}[c]{0.32\textwidth}
         \centering
         \includegraphics[width=\textwidth]{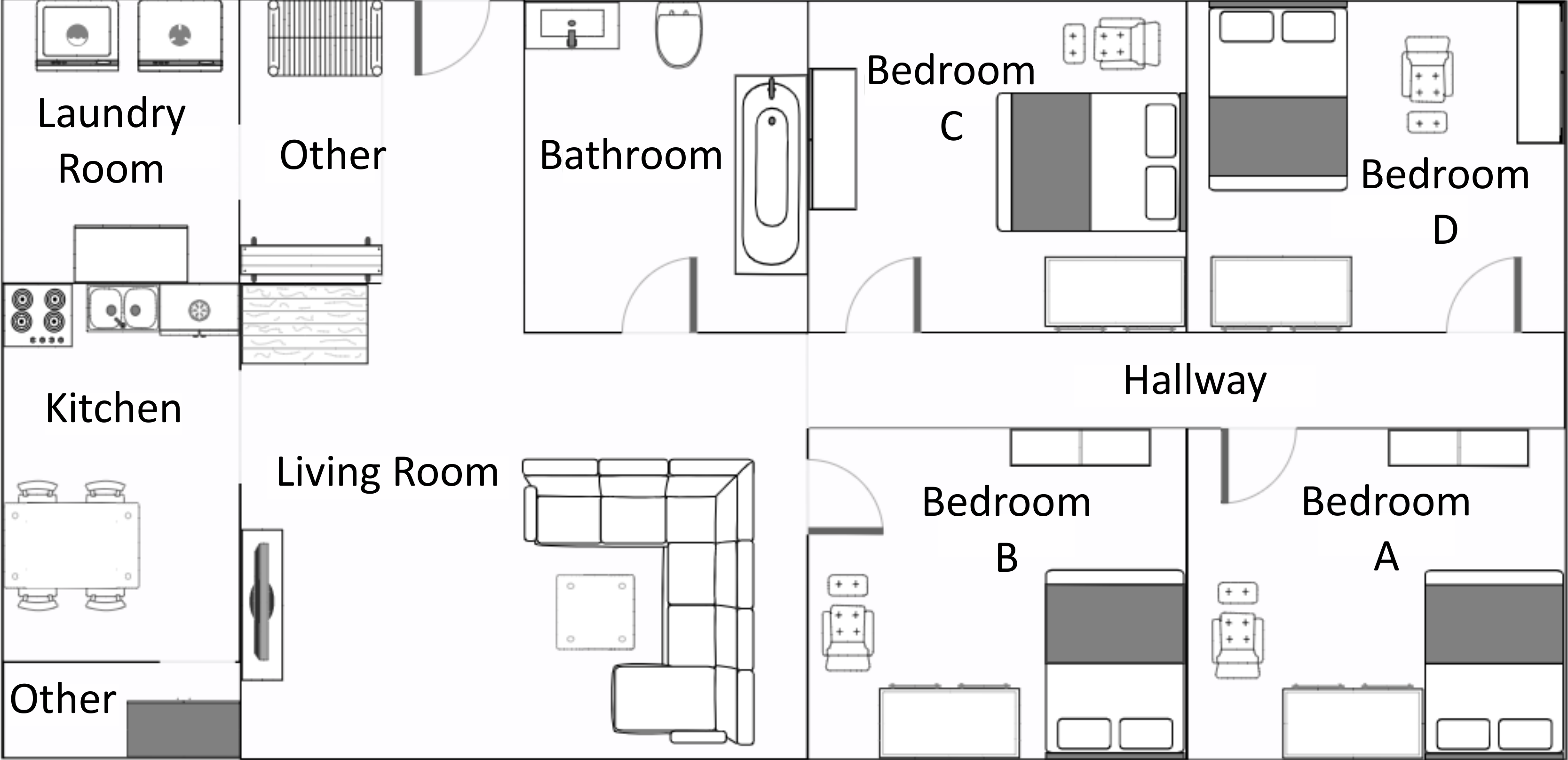}
         \caption{Floorplan}
         \label{fig:floorplan}
     \end{subfigure}
     \hfill
     \begin{subfigure}[c]{0.32\textwidth}
         \centering
         \includegraphics[width=\textwidth]{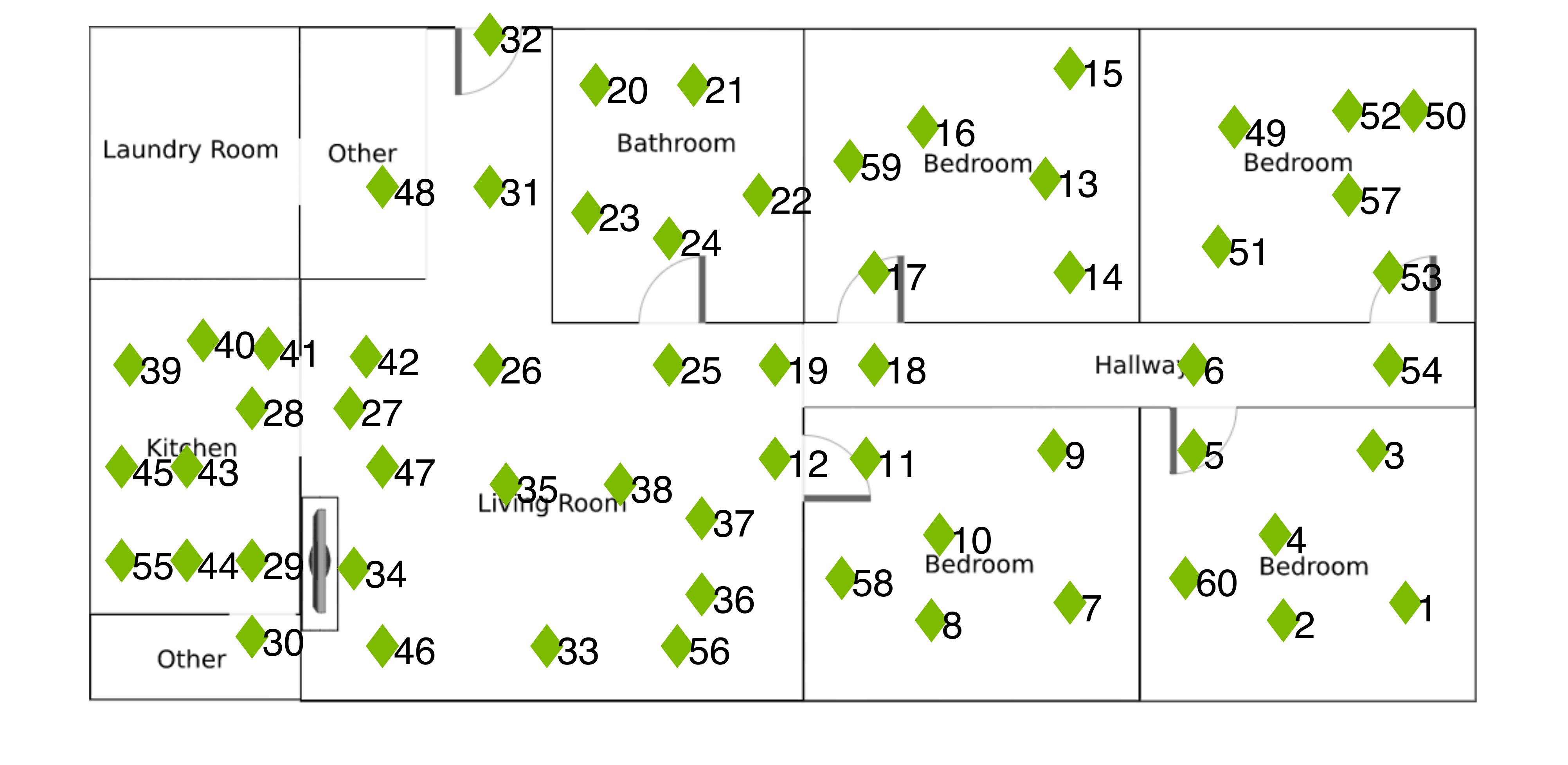}
         \caption{Sensor layout}
         \label{fig:senlay}
     \end{subfigure}
     
        \caption{The experimental space for the four-person scenario.
        Other multi-occupancy scenarios have similar settings. 
        }
        \label{fig:sensorlayout}
\end{figure*}

Data associating problem (i.e., identification annotation) is one of the central problems for the sensor-based activity recognition in multi-occupancy smart homes~\cite{benmansour2015multioccupant, chiang2010interaction, leal2016learning, prossegger2014multi}. 
It refers to labeling the identifications of the time-series sensor events by mapping each sensor event with the resident causing its generation.
A suitable model of data associating is the prerequisite for leveraging single-occupancy techniques into multi-occupancy households.
Current solutions on the identification annotation are mainly relying on self-reporting~\cite{tapia2004activity} and camera-recording~\cite{adib2015multi}, where the former one is biased and the latter one would invigilate the privacy. 
The capability for automatically labeling the identification, hence, is significant for a practical smart home system.
Researchers tend to use wearable sensors to reduce the complexity of the problem because wearable sensors can be utilized as the identification tag of different residents~\cite{mokhtari2018multi}.
Another promising way is to leverage the real-time locating system (RTLS) to locate different residents when they are interacting with the environment. 
RTLS is a rising technology for detecting both the location and identification of the target, where the target could refer to an item, a person, or a vehicle~\cite{boulos2012real, zafari2019survey}. 
By integrating the residents' location and sensor layout, we annotate sensor events with identifications using the RTLS-based approach.

In this paper, taking the automatic annotation problem as the desired context-aware task to solve, we explore the interaction dynamics between the sensor-based environment and multiple residents by proposing the \emph{MoSen} emulation environment.
\mosen~is designed to evaluate the way in which different localization resolutions, number of residents, number of sensors, and varying sensor deployments affect the performance of the pre-designed sensor network before real deployment.
By investigating the dynamics of annotation accuracy, given any floorplan, \mosen~is able to provide sensor selection strategy and metric-based design suggestion for the pre-designed sensor layout. 
\mosen~can help practitioners with cost-effective and accurate sensor integration for any given home deployment.

By using real behavior models and synthetic data, we emulate multi-occupancy scenarios in households with 2 to 5 residents.
Given the scarcity of multi-occupancy dataset and difficulty in realistic data collection with existing technologies, especially in the circumstance of the \textit{COVID-19} pandemic with social distancing, we provide a novel framework to generate synthetic multi-occupancy behavior models by modeling real single-occupancy datasets that collected in real homes.
The quality of the multi-occupancy behavior model is validated by comparing the performance between synthetic and real double-occupancy datasets.

The main objective of this paper is to offer an effective evaluation structure and feasible sensor selection strategy to \emph{any} smart home. Through comparing real and synthetic datasets, we discuss potential challenges when sensor-based activity recognition is adopted in different multi-occupancy scenarios.
Our main contributions of the paper are as follows:
\begin{itemize}
  \item We propose \emph{MoSen} to investigate the interaction dynamics between sensor-based environment and multiple residents;
  
  \item We provide an algorithm to generate synthetic multi-occupancy behavior models and compare the performance with the real dataset;
  
  \item We explore how the labeling accuracy is affected by different localization resolution, the residents' quantity, sensor density, and varying deployment in multi-occupancy scenarios;
  
  \item We design a sensor selection strategy to balance the trade-off between deployment costs and expected labeling accuracy in different homes, which accelerates the practical adoption of sensor-based activity recognition in reality.
\end{itemize}

The rest of this paper is organized as follows. 
In Section~\ref{sec:related} we present the related work, 
Section~\ref{sec:overview} represents an overview of \mosen~system, and Section~\ref{sec:method} describes the design methodologies applied in the proposed system. In Section~\ref{sec:evluation}, we evaluate the effect of localization devices resolutions, residents' quantity and sensor density, respectively. With the analytical result, we provide a case study in Section~\ref{sec:strategy}, then present discussions in Section~\ref{sec:discussion}, and the final conclusion in Section~\ref{sec:conclusion}.

\section{Related Work}
\label{sec:related}

\subsection{Activity Recognition}

Human indoor activities are complex, diverse and stochastic, making them challenging to define and quantify.
A variety of advanced ubiquitous sensing technologies (\eg~ wireless sensing~\cite{karanam2019tracking}, wearable sensors~\cite{bianchi2019iot}, or ambient sensors~\cite{cook2013transfer}) have been adopted to collect human indoor activity data~\cite{abbas2019wideep, adib2015smart, alemdar2013aras, cook2012casas}. 
Human activity recognition is a central task for accelerating automation integration in smart environments~\cite{rodriguez2020context}.
Prior works have illustrated modeling human activity patterns is valuable for providing personalized services~\cite{sztyler2017online} or context-aware interactions with the resident~\cite{cao2018gchar, lee2015comon+}. 
Currently, human activity can be recognized using two main approaches, namely video-based~\cite{jalal2017robust} and sensor-based~\cite{cook2013transfer}.
In this paper, we focus on sensor-based activity recognition.

\newcommand{\tabincell}[2]{\begin{tabular}{@{}#1@{}}#2\end{tabular}}  
\begin{table*}[t!]\scriptsize
  \caption{Comparison of main indoor positioning technologies (I) 
  }
  \label{tab:compareI}
    \begin{tabular}{cccccccccc}
    \toprule
             & Infrared (IR) & Ultrasound & Acoustic Signal & Visible Light (VLC) & UWB-based & RFID-based \\
    \midrule
            Positioning Accuracy & 0.57 - 2.3m~\cite{brena2017evolution} 
            & 10mm~\cite{schweinzer2010losnus} & 5.4cm~\cite{moutinho2016indoor} - meters~\cite{rishabh2012indoor} 
            & 1mm~\cite{nadeem2014highly} - 45cm~\cite{afzalan2019indoor} & 10cm~\cite{cheng2019uwb} - 50cm~\cite{ruiz2017comparing} & 15cm~\cite{xu2017rfid} - meters~\cite{ni2003landmarc}\\
    \midrule
            Coverage Range & Sub-room & Room & Room & Room & Room & Sub-room\\
    
    \midrule
            Cost & Low & High & Medium & Low  & High & Medium\\
    \midrule
            \tabincell{c}{Infrastructure Complexity} & High & Medium & Medium & Low & High & Medium\\
    \midrule
            Network 
            & \tabincell{c}{IR sensor network \\ + IR tags} 
            & \tabincell{c}{Activation unit \\ + 6 broadband \\ US transmitters \\ (Polaroid 600)~\cite{brena2017evolution}} 
            & \tabincell{c}{Virtex 5 FPGA-based board \\ + 4 speakers~\cite{aguilera2018broadband}} 
            & \tabincell{c}{5 LED lights \\ + 1 receiver~\cite{nadeem2014highly}} & \tabincell{c}{DW1000 UWB ranging chip, \\ processor STM32F105 \\ ARM Cortex M3, \\ omnidirectional antenna, \\ nodes or tags~\cite{ruiz2017comparing}} 
            & \tabincell{c}{RFID reader, \\ reference tags, \\ target tags~\cite{xu2017rfid}}\\
    \midrule
            Localization Method & \tabincell{c}{Proximity detection, \\ Trilateration} & Trilateration & \tabincell{c}{Watermarking, \\ Trilateration} & \tabincell{c}{Lateration, \\  Angulation, \\ Fingerprinting} & \tabincell{c}{Multilateration} & \tabincell{c}{Triangulation, \\Fingerprinting, \\ Proximity detection}\\
    \midrule
            \tabincell{c}{Frequently-used \\ Convention \\ Measurement} & ToA & RSS, ToF & \tabincell{c}{ToF, TDoF \\ Phase conherence \\ ToA, TDoA} & \tabincell{c}{RSS, TDoA, AoA} & \tabincell{c}{ToF, ToA, TDoA} & \tabincell{c}{RSS, ToA, AoA}\\
    \midrule
            \tabincell{c}{Multi-person \\ Scenario} 
            & \tabincell{c}{7m x 7m area, \\ with 4 PIR sensors \\ on the 4 corners, \\ locate 3 persons \\ within 1.25 meters~\cite{yang2020deep}} 
            & \tabincell{c}{4m x 4m area, \\ with 16 receivers, \\ locates more than \\ 70 separate transmitters \\ within 3cm~\cite{brena2017evolution}}
            & \tabincell{c}{3m x 3m area, \\ with 4 speakers, \\ locates persons \\ within 20 cm~\cite{aguilera2018broadband}} & \tabincell{c}{3m x 3m area, \\ with 4 LEDs, \\ locates receivers \\ within 33cm~\cite{liang2017plugo}}
            & \tabincell{c}{5m x 7m area, \\ locates 3 persons \\ within 11.7cm~\cite{adib2015multi}}
            & \tabincell{c}{3.6m x 4.8m area, \\ with 117 reference tags, \\ locates 5 targets tags \\ within 15cm~\cite{xu2017rfid}}\\
            
    
    \bottomrule
    \end{tabular}
    \scriptsize{\tabincell{c}{VLC: Visibale Light Communication; UWB: Ultra-Wideband; RFID: Radio Frequency Identification; ToA: Time of Arrival; \\ TDoA: Time Difference of Arrival ; RSS: Received signal strength; ToF: Time of Flight; TDoF: Time Difference of Flight; AoA: Angle of Arrival;}}
\end{table*}

\subsubsection{Sensor-based Activity Recognition}

Sensor-based activity recognition utilizes sensor readings to understand human activities.
The metadata emanated from multifarious sensors embedded in the living environment. 
These data will be trained and learned by a series of machine-learning or deep-learning algorithms~\cite{wang2019deep}.
In this paper, we leverage human activity datasets from real homes that deployed with different ambient sensors (\ie~motion sensor, temperature sensor, light sensor)~\cite{alemdar2013aras, cook2012casas}.

\subsubsection{Multi-person Activity Datasets}

The majority of researches in human activity recognition have investigated the \emph{single-occupancy scenario}~\cite{van2008accurate, prossegger2014multi}, where only one resident lives in a single space. 
However, the real environment is usually inhabited by more than one resident and even with pets, which is referred to as \emph{multi-occupancy scenario}~\cite{alemdar2013aras} in this paper. 
Multi-person activity recognition has less investigation, as many practical challenges are yet to be overcome in the single-occupancy scenario~\cite{benmansour2015multioccupant}. 
Recent pilot deployments demonstrate the applicability and adaptability of multi-occupancy scenarios by using different machine learning algorithms~\cite{benmansour2015multioccupant,adib2015multi,mokhtari2018multi}. 
There are two publicly and widely-used multi-person datasets in current literature, the \emph{CASAS Datasets}~\cite{cook2012casas}, and the \emph{ARAS Datasets}~\cite{alemdar2013aras}.
We compare our synthetic multi-person behavior models with these two real datasets to validate the quality of the synthetic model.

\subsection{Real-time Locating System}

Real-time locating system (RTLS) is a rising technology for detecting both the location and identification of the target, where the target could refer to an item, a person, or a vehicle~\cite{boulos2012real, zafari2019survey}. 
Different positioning technologies have been investigated in the last several decades, while these technologies perform a similar task with varying accuracy.
We conclude 12 main indoor positioning technologies in Table~\ref{tab:compareI} and Table~\ref{tab:compareII}, comparing the \textit{positioning accuracy, coverage range, cost, infrastructure complexity, network, localization method, and frequently-used convention measurement} of different technologies.

Applications for RTLS, also called location-based services (LBSs), have already been broadly adopted in a variety of indoor location-aware scenario~\cite{lenders2008location, uttama2015loced}, from mapping and navigation services~\cite{bendavid2013rfid,moreira2015wi} to human-robotics interaction~\cite{boulos2012real}. 
In the transition from single-occupancy scenario to the multi-occupancy environment, it becomes significantly important to track each resident~\cite{crandall2013tracking}. 
In Table~\ref{tab:compareI} and Table~\ref{tab:compareII}, we also conclude how 12 main indoor positioning technologies perform in multi-occupancy environment, as listing their basic experimental setting and locating accuracy.
With tracking residents respectively in an efficient and accurate approach, the sensor events can be separated into different steams, as each resident would have an independent data-driven profile that serves for further personalized interaction provided by the smart environment.

\subsubsection{Trajectory with identification}
\label{subsubsec:tra_iden}

Recent works have shown the capability by tracking resident's trajectory with their identification by non-camera-based systems~\cite{brena2017evolution,aguilera2018broadband,nadeem2014highly,ruiz2017comparing,xu2017rfid,pan2015indoor}. 
These classes of technologies can be categorized into \emph{device-based} system and \emph{device-free} system. 
Device-based system use smartphones, smartwatches, or other wireless tags embedded into the human body. 
These extra devices will be leveraged to identify different individuals~\cite{lane2010survey,xu2017rfid,abbas2019wideep,komai2016beacon,bocca2013multiple}. 
Device-free system depends on wireless signals by analyzing the signal patterns from the breathing or heartbeat to perform identification~\cite{adib2015multi,adib2015smart,mshali2018survey}.

\begin{table*}[t!]\scriptsize
  \caption{Comparison of main indoor positioning technologies (II)
  }
  \label{tab:compareII}
    \begin{tabular}{cccccccccc}
    \toprule
             & WLAN-based & Bluetooth & Zigbee & Vision & Geomagetism  & Inertial Navigation (INS) \\
    \midrule
            Positioning Accuracy & 23cm~\cite{xiong2013arraytrack} - 5m~\cite{ruiz2017comparing} 
            & 2m - 10m~\cite{zafari2019survey}
            & 25cm - 5m & 1cm~\cite{mulloni2009indoor} - 2m~\cite{morar2020comprehensive} & 1m -5m~\cite{he2017geomagnetism} & 1m - 10m~\cite{mendoza2019meta}\\
    \midrule
            Coverage Range & Multiple-Room & Multiple-Room & Multiple-Room & Sub-room & Building & Building \\
    \midrule
            Cost & Medium & Medium & Medium & High & Low & Low \\
    \midrule
            \tabincell{c}{Infrastructure Complexity} & Low & Low & Low & High & Low & Low\\
    \midrule
            Network & Wi-Fi APs + Phones~\cite{abbas2019wideep} & Beacon network + BLE Tags~\cite{komai2016beacon} & 32 Zigbee nodes~\cite{bocca2013multiple}
            & Cameras~\cite{morar2020comprehensive}
            & Magnetometer~\cite{ma2016basmag}
            & Inertial sensors~\cite{kang2014smartpdr}\\
    \midrule
            Localization Method & \tabincell{c}{Fingerprint, \\ Trilateration} 
            & \tabincell{c}{Proximity detection, \\ Multilateration, \\ Centroid determination}
            & \tabincell{c}{Proximity detection, \\ Multilateration} & \tabincell{c}{Computer \\ Vision} & Fingerprint & Pedestrian Dead Reckoning \\
    \midrule
            \tabincell{c}{Frequently-used \\ Convention \\ Measurement} & RSSI, CSI, ToF, AoA, TDoA & RSSI, AoA, ToF & RSSI, AoA, ToF & \tabincell{c}{Traditional \\ Image Analysis, \\ AI} & RSS &  Inertial Measurement unit \\
    \midrule
            \tabincell{c}{Multi-person \\ Scenario} & \tabincell{c}{3m x 3m area, \\ with 4 Wi-Fi NICs, \\ locates 3 person \\ within 55 cm~\cite{karanam2019tracking} }
            & \tabincell{c}{6.1m x 9.4m area, \\ with 6 beacons \\ in six rooms, \\ locates 2 persons \\ within 2 meters~\cite{komai2016beacon}}
            & \tabincell{c}{7m x 8.25m area, \\ with 32 sensors, \\ locates 5 persons \\ within 55cm~\cite{bocca2013multiple}}
            & \tabincell{c}{2.2m x 6m area, \\ with two cameras, \\ locates objects \\ within 7.1cm~\cite{shim2015mobile}}
            & \tabincell{c}{58m x 42m area, \\ locates persons \\ within 2 meters~\cite{ma2016basmag}}
            & \tabincell{c}{Locates 5 persons \\within 1.62 meters~\cite{kang2014smartpdr}}
            \\
    
    \bottomrule
    \end{tabular}
    \scriptsize{\tabincell{c}{WLAN: Wireless Local Area Network; INS: Inertial Navigation System; BLE: Bluetooth Low Energy; RSSI: Received Signal Strength Indicator; CSI: Channnel State Information; AI: Artificial Intelligence}}
\end{table*}

\subsubsection{Trajectory and Activity Recognition}

The real-world experiment conducted by~\citeauthor{nguyen2005learning}~\cite{nguyen2005learning} emphasized the applicability in modeling complex activities from human indoor trajectories. 
Their work also demonstrates the feasibility in recognizing activities from new trajectories~\cite{nguyen2005learning} by applying the hierarchical hidden Markov model (HHMM).
\citeauthor{wilson2005simultaneous}~\cite{wilson2005simultaneous} have demonstrated that localization accuracy and activity recognition can be beneficial to each other, especially in multi-occupancy environments. 
\citeauthor{lu2009robust}\cite{lu2009robust} provide more fine-grained outcomes in a single-occupancy scenario to illustrate the possibility of location-aware activity recognition.
These works on location-based activity recognition validate the feasibility of leveraging residents' locations to annotate the time-series sensor events.

\subsection{Significance of Data Annotation}

Recent advances in machine learning and deep learning accelerate the sensor-based activity recognition but most of them require annotated datasets~\cite{wang2019deep}.
The quality of label extends a significant impact on the performance of machine learning models.
However, collecting sensor data with ground-truth labels (\ie, identification, activity) is still challenging, especially in longitudinal monitoring scenarios. 
Currently, ground-truth labels are obtained either from a resident's diary~\cite{intille2003context} or video-based recording techniques~\cite{adib2015multi}.
In order to simplify the annotation process, some researchers also designed a simple graphical user interface (GUI) to help residents finish the diary report~\cite{alemdar2013aras}.
However, this can be tedious, time-consuming, and inaccurate.
Unlike the diary-based technique, the video-based recording is unobtrusive and precise, but incorporates major privacy concerns and needs extra manual inputs. 
The capability of automatic annotation, hence, is the central primitive when building each resident's individual activity profile.
\citeauthor{hamm2012automatic}~\cite{hamm2012automatic} have presented a flexible framework on combining heterogeneous sensory modalities with classifiers for sequence labeling automatically.

In this paper, we are interested in the identification labeling for time-series sensor events and leave activity labeling to a future research.
Previous works (as listed in Section~\ref{subsubsec:tra_iden}) have already demonstrated the capability of a trajectory to identify persons.
Hence, in our proposed \mosen~system, we leverage these identified trajectories to annotate sensor events automatically in real-time, by integrating residents' respective locations and the sensor layout.
For instance, in a 4-person scenario, where there are 4 residents in the home, the location information (refers to four respective location points) and the sensor layout are known, the proximity between each location point and the triggered sensor will be compared and the nearest location points from that sensor would be selected.
Such a solution mainly depends on the accuracy of the localization techniques (as details are shown in Table~\ref{tab:compareI} and Table~\ref{tab:compareII}) and ambient-sensor density.
We discuss the effects of different localization resolutions in different sensor layout in the Section~\ref{sec:evluation}.

\subsection{Synthetic Sensor Data Generation}

In order to protect users' privacy and increase data sharing, synthetic data generation has been developed as an alternative tool among data scientists~\cite{jordon2018pate, dwork2008differential}. 
The generated data preserves the required statistical features as the real data in a non-adversarial setting, and is hardly distinguishable from the real data when the generation structure is mandated by an adversarial network~\cite{bellovin2019privacy}.
Effectively generating synthetic data can augment the labeled data and compensate for the data scarcity, when the availability of labeled data is constrained~\cite{masud2012facing}.
Especially for sensor-based activity data, where data collection for even single-occupancy scenario is hard, there are more challenges posed in multiple occupancy scenarios.
The aforementioned gap demonstrates the importance of synthetic sensor-data generation in our multi-person setting.

Recent works on Generative Adversarial Networks (GAN) have demonstrated their capability in generating different types of data, from images generation~\cite{brock2018large,frid2018gan}, text generation~\cite{zhang2017adversarial,xu2018attngan}, music composition~\cite{kulkarni2019survey}, and time-series sensory data generation~\cite{alzantot2017sensegen}.
The research published by~\cite{dahmen2019synsys} employed hidden Markov models (HMMs) to generate realistic synthetic smart home sensor data. The authors used data similarity measures to validate the realism of generated data, which are not random but preserves the underlying patterns or structures of the real data. 
In this paper, we compare the performance of our synthetic multi-person behavior models with the real datasets to validate the effectiveness of our synthetic models.


\section{System Overview}
\label{sec:overview}

\begin{figure}[t]
  \centering
  \includegraphics[width=3.3in]{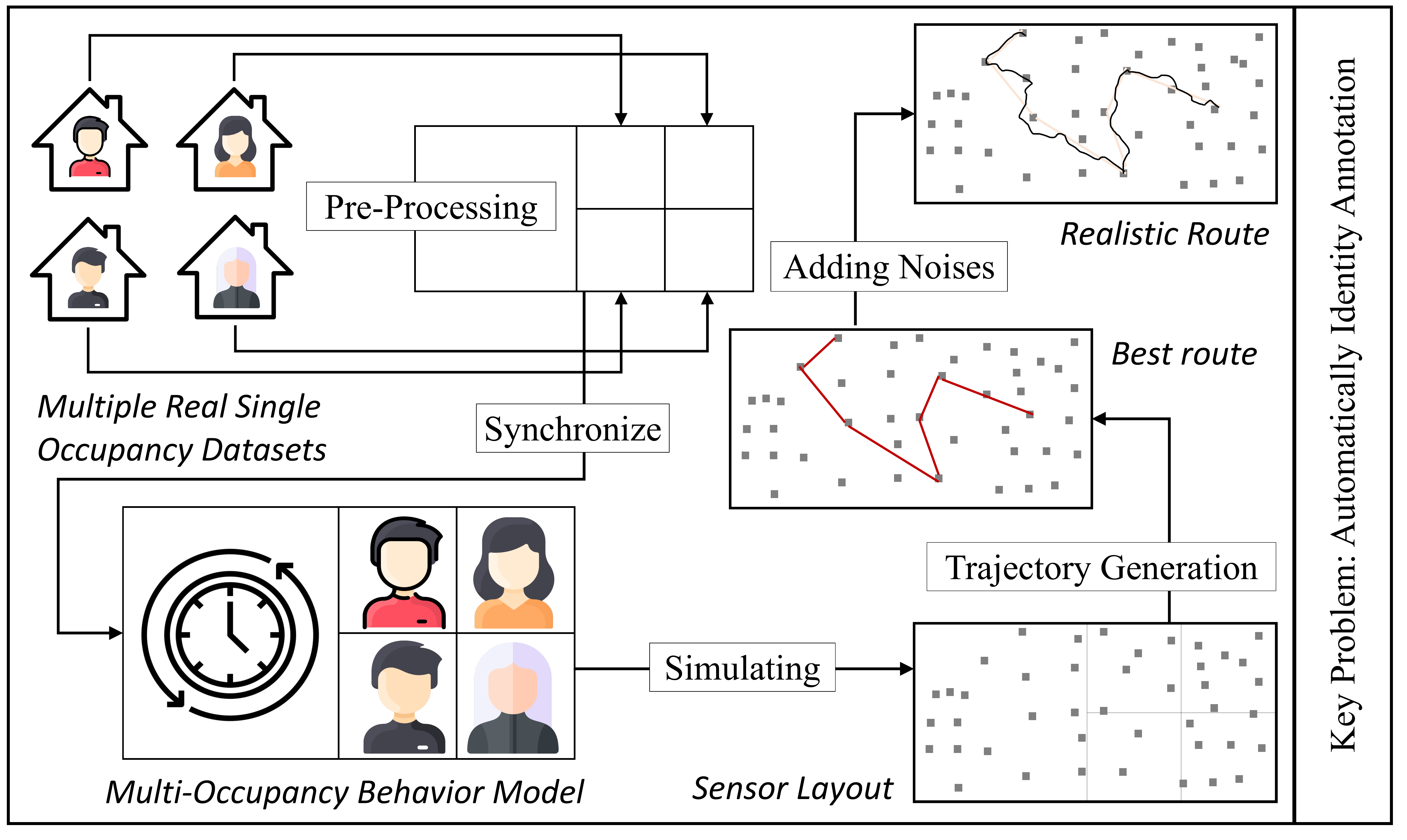}
  \caption{An overview of the \mosen~system
  }
  \label{fig:structure}
  \Description{Structure.}
\end{figure}

The design of \mosen~system is motivated by a need to accelerate the practical implementation of sensor-based activity recognition technologies in \emph{multi-occupancy} settings.
\mosen~is adaptable to \emph{any} customized indoor environments.
Smart home designers or practitioners can leverage the analytical result of their pre-design sensor network in the specific context-aware task to better balance the trade-off between the deployment cost and system performance.

Figure~\ref{fig:structure} shows an overview of \mosen~system in solving the identification annotation problem, which is one of the central problems for the sensor-based activity recognition in multi-occupancy smart homes~\cite{benmansour2015multioccupant}.
It refers to labeling the identifications of the time-series sensor events by mapping each sensor event with the resident causing its generation.
In this paper, we solve the identification annotation problem by leveraging the \textit{Graph and Rule-Based Algorithm (GR/ED)} proposed in \cite{crandall2013tracking}, which was designed to track individuals in an ambient sensor setting. The core idea is that individuals trip sensors when they move from one location to another. 
Sensor events will then be separated into different streams by leveraging human trajectory or location information with the \textit{nearest neighbor standard filter (NNSF)}~\cite{bar1990tracking}, a classical data association method.
To achieve the required labeling accuracy of identifications for time-series sensor events in the multi-occupancy environment, \mosen~can additionally provide a \emph{sensor selection strategy} that fits for the user's requirements while optimizing for the number of sensors and their placement (hence the installation cost) to achieve the highest labeling accuracy.

\mosen~platform can emulate this annotation process with different sensor settings for any pre-designed smart home.
The platform is assumed that only the sensor events are provided by the sensors, and without the need to considering how heterogeneous or multi-modal sensing environments are meshed and combined. 
In a practical setting, the sensor event is recorded when a sensor is triggered. In \mosen, we emulate triggering sensors by building realistic single-person activity patterns. 
In this way, we can add different representative activity patterns into \mosen~to simulate multi-resident scenarios, noted as the \textit{multi-occupancy behavior model} in Figure~\ref{fig:structure}.
Due to the stochastic nature of our choices and the heterogeneity of chosen single-person datasets, residents that contribute to each activity pattern might have different backgrounds, different habits, and different daily routines. 

\begin{table}[t!]\footnotesize
  \centering
  \caption{Details of the five CASAS single-occupancy testbeds}
  \footnotesize{"D" indicates magnetic door sensors; "L" indicates light switches; \\ "LS" indicates light sensors; "M" indicates infrared motion sensors; \\ "MA" indicates wide-area infrared motion sensors; "T" indicates temperature sensors}\\
  \label{tab:fivebed}
    \begin{tabular}{ccccccccc}
    \midrule
             \multirow{2}{*}{Testbed} & 
             \multirow{2}{*}{Timespan} &  
             \multicolumn{7}{c}{Number of Sensors}   \\ 
             \cline{3-9}
             &  & D & L & LS & M & MA & T & Total \\
    \midrule
            hh120 & \tabincell{c}{2012.1.28-3.31} & 3 & 7 & 15 & 11 & 4 & 4 & 44\\
    \midrule
            hh122 & \tabincell{c}{2013.4.1-4.30} & 4 & 0 & 24 & 19 & 5 & 5 & 57 \\
    \midrule
            hh123 & \tabincell{c}{2013.3.2-4.1} & 2 & 0 & 14 & 4 & 10 & 6 & 36 \\
    \midrule
            hh125 & \tabincell{c}{2013.3.1-4.10} & 2 & 0 & 0 & 15 & 0 & 3 & 20 \\
    \midrule
            hh126 & \tabincell{c}{2013.8.1-9.6} & 0 & 0 & 0 & 15 & 0 & 0 & 15 \\
    \midrule
    \end{tabular} \\
    
\end{table}

\begin{figure}[t!]
  \centering
  \includegraphics[width=3.4in]{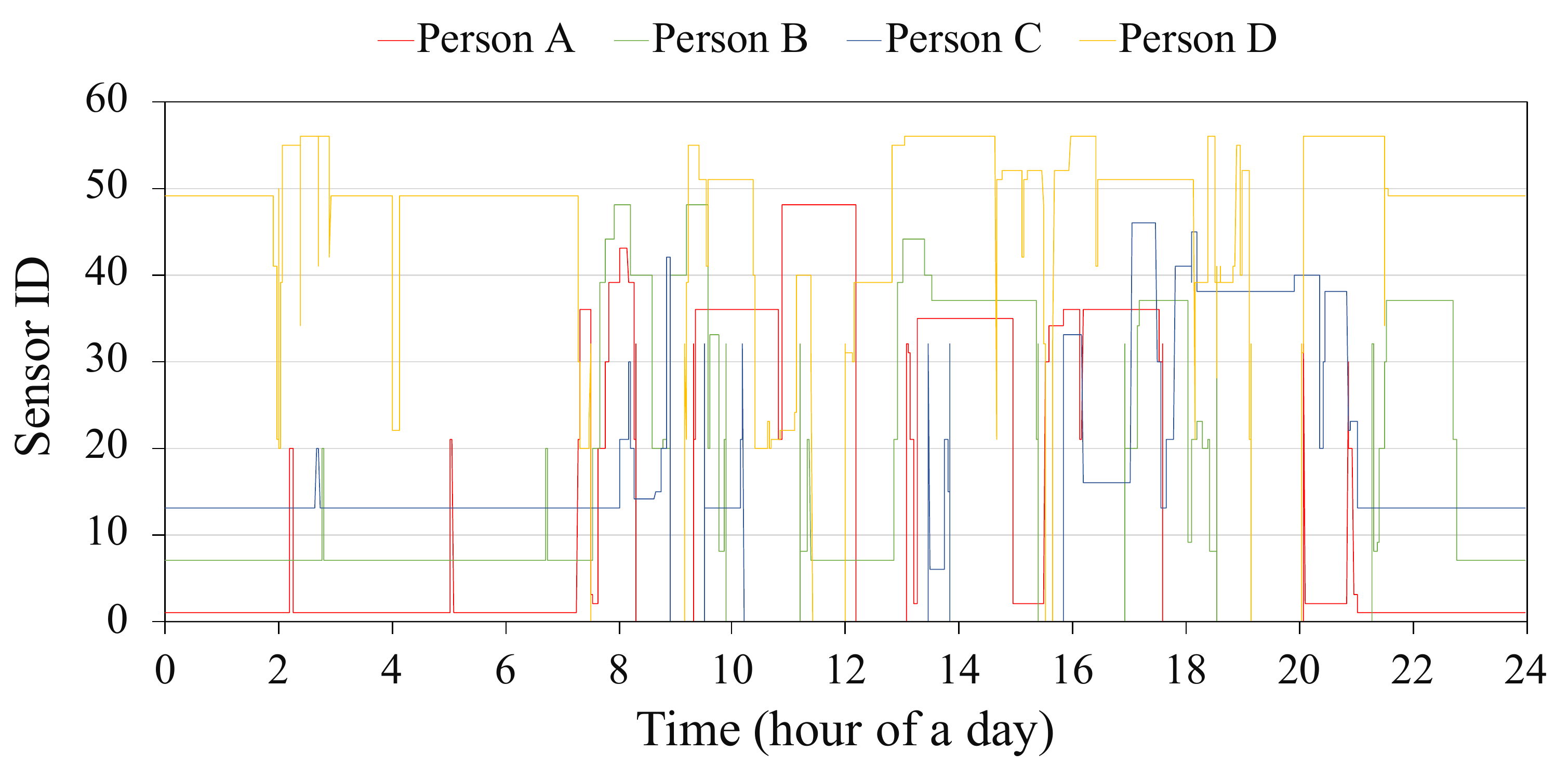}
  \caption{Overview of sensor activation list in a four-person scenario
  }
  \label{fig:sensorlist}
  \Description{Sensorlist.}
\end{figure}


We then leverage \textit{Dijkstra' algorithms}~\cite{dijkstra1959note} to emulate and generate each resident's daily trajectory.
These trajectories are utilized as \textit{ground-truth trajectories} of residents.
Normally distributed noise, depending on the different resolutions of positioning technologies, is added to the ground-truth value to generate a new trajectory that emulates how localization devices work.
This noise-added trajectory is referred to as the detected trajectory. 
\textit{Labeling accuracy}, in this paper, is defined as how \emph{detected trajectory} from different sensor networks affect the identification annotation process.

With \mosen, by combing the real residents' activity pattern and floorplan, every multi-occupancy house can be emulated and evaluated before deploying a real sensor system, which we believe can accelerate the practical utility of sensor-based activity recognition system in smart homes.

\section{System Design Methodology}
\label{sec:method}

\begin{table}[t!]\footnotesize
  \caption{Correlation between real single-occupancy testbeds and synthetic multi-occupancy behavior model
  }
  \label{tab:realsyn}
    \begin{tabular}{ccl}
    \midrule
            Multi-occupancy Model & Space Size & Relative Single-occupancy Dataset \\
    \midrule
            2-Person Scenario  & 12.5m x 8m & hh120, hh122 \\
    \midrule
            3-Person Scenario & 16.5m x 8m & hh120, hh122, hh123 \\
    \midrule
            4-Person Scenario & 16.5m x 8m & hh120, hh122, hh123, hh125\\
    \midrule
            5-Person Scenario & 20.5m x 8m & hh120, hh122, hh123, hh125, hh126  \\
    \midrule
    \end{tabular} \\
\end{table}

\subsection{Dataset Description}

In the sensor-based activity recognition, multi-modal sensor readings are collected and then represented as time-series data to describe human indoor activities.
The dataset contains a series of sensor events ordered by time.
Each sensor event is recorded when the respective sensor is triggered or activated, at the time when this sensor is touched or walked around by residents.
In this paper, we choose two widely-used published datasets in our analysis, CASAS datasets~\cite{cook2012casas} and ARAS datasets~\cite{alemdar2013aras}. 

\subsubsection{CASAS datasets}
The CASAS group collected human activity datasets from the WSU smart apartment test bed~\cite{cook2012casas}.
Activity labels are annotated in CASAS datasets with respective start time and end time, via a handwritten diary.
The majority of datasets represent indoor activities as a series of sensor events, which contains the event timestamp, the sensor name, the sensor state and the activity label. 
Each sensor event should at least contains the following details:

\textit{[Timestamp, Sensor ID, Sensor Status, Activity Label]}

In this paper, five single-occupancy testbeds chosen from CASAS datasets are leveraged to generate the synthetic multi-person behavior model. These five testbeds are annotated as hh120, hh122, hh123, hh125, hh126~\cite{cook2013transfer}. 
Details and properties of them are shown in Table~\ref{tab:fivebed}.
And the relations between these testbeds and synthetic multi-occupancy datasets are demonstrated in the Table~\ref{tab:realsyn}.
Activity labels contained in single-occupancy testbeds include:
\textit{bed-toilet transition, cook, eat, enter home, leave home, personal hygiene, phone, relax, sleep, work.}

\subsubsection{ARAS datasets}

Different from the CASAS group collecting the activity in the lab setting, the ARAS group collected two pairs of residents' daily activities in their real houses by recording the ground truth labels with a designed Graphical User Interface (GUI)~\cite{alemdar2013aras}. For each house, it consists of 30 days sensor reading as a form of 22x86400 matrix for each day, where the first 20 columns (\emph{S1 - S20}) refer to the binary sensor reading and the column 21 (\emph{P1}) and 22 (\emph{P2}) are the activity labels for resident A and B. 
The activity labels, ranging from 1 to 27, represent 27 different activities.
They are in order as follows:
\\
\textit{going out, preparing breakfast, having breakfast, preparing lunch, having lunch, preparing dinner, having dinner, washing dishes, having snack, sleeping, watching TV, studying, having shower, toileting, napping, using internet, reading book, laundry, shaving, brushing teeth, talking on the phone, listening to music, cleaning, having conversation, having guest, changing clothes, others.}

\normalsize

The data example below presents how ARAS dataset represent sensor status at every second and the activity labels of two residents.

\begin{table}[h]\scriptsize
    \setlength{\tabcolsep}{0.4mm}{
    \begin{tabular}{ccccccccccccccccccccccc}
             \textit{Timestamp} & \textit{S1} & \textit{S2} & \textit{S3} & \textit{S4} & \textit{S5} & \textit{S6} & \textit{S7} & \textit{S8} & \textit{S9} & \textit{S10} & \textit{S11} & \textit{S12} & \textit{S13} & \textit{S14} & \textit{S15} & \textit{S16} & \textit{S17} & \textit{S18} & \textit{S19} & \textit{S20} & \textit{P1} & \textit{P2}  \\
    \midrule
            86398 & 0 & 0 & 1 & 0 & 0 & 0 & 0 & 0 & 0 & 0 & 0 & 0 & 0 & 0 & 0 & 0 & 0 & 0 & 0 & 0 & 12 & 2 \\
    \midrule
            86399 & 0 & 0 & 0 & 0 & 0 & 0 & 0 & 1 & 0 & 0 & 0 & 0 & 0 & 0 & 0 & 0 & 0 & 0 & 0 & 0 & 3 & 2 \\
    \midrule
    \end{tabular}} \\
    \footnotesize{\textit{S1 to S20 indicate the sensor identification, and their statuses are shown in binary form, where 0 refers deactivated status and 1 refers to activated status; P1 and P2 indicate the activity label for Person A and Person B, respectively. For instance, 12 represents "watching TV", 2 represents "going out".}}
\end{table}

\subsubsection{Format of Synthetic Datasets}
\label{subsubsec:format_data}

Two aforementioned datasets~\cite{cook2013transfer, alemdar2013aras} illustrate two main variations of data representations in the sensor-based activity recognition literature. 
In this paper, further analysis of both datasets requires us to integrate them in a uniform way.
By this motivation, we define the format of our synthetic dataset to include the critical information we need, as the sample data are shown as followed:

\begin{table}[h]\tiny
    \centering
    \setlength{\tabcolsep}{2mm}{
    \begin{tabular}{cccccc}
             \textit{Timestamp} & \textit{P1} & \textit{P2} & \textit{P3} & \textit{P4} & \textit{P5}  \\
    \midrule
            t$_1$ & S$_1^1$ & S$_2^1$ & S$_3^1$ & S$_4^1$ & S$_5^1$ \\
    \midrule
            t$_2$ & S$_1^2$ & S$_2^2$ & S$_3^2$ & S$_4^2$ & S$_5^2$ \\
    \midrule
            ... & ... & ... & ... & ... & ... \\
    \midrule
            t$_i$ & S$_1^i$ & S$_2^i$ & S$_3^i$ & S$_4^i$ & S$_5^i$ \\
    \midrule
    \end{tabular}} \\
\end{table}

\textit{Timestamp} refers to \textit{critical timestamps} in multi-person scenario, where \textit{critical} indicates that there are at least one sensor triggered in that second. Every critical sensor activation also noted as one \textit{sensor event} in this paper.
\textit{P1 to P5} represent five residents, respectively.
S$_n^i$ denotes the sensor ID triggered at t$_i$ by the resident \textit{Pn}.
We leverage sensor activation that annotated with effective activity labels from CASAS and ARAS dataset to model resident activity patterns, which are action-based behavioral models.

\subsection{Data Pre-processing}

Data pre-processing is responsible for unifying the format of the dataset from different sources.
We pre-process the public single-occupancy datasets, by developing an algorithm to capture the critical timestamps for sensor status transition or activity transition, then format the data as discussed aforementioned in Section~\ref{subsubsec:format_data}.

\subsubsection{Capturing Critical Timestamps}
In CASAS datasets~\cite{cook2013transfer}, sensor events are recorded in order with timestamps that sensors are triggered. 
However, for the ARAS dataset~\cite{alemdar2013aras}, data is recorded in every second, that is, the dataset has 86399 lines which refers a day has 86399 seconds. 
Capturing critical timestamps for both datasets is the first step to uniform the data.

\subsubsection{Defining the Start and End Point}
Critical timestamps are important for us to know how the sensor events or activities transit. In identifying the start or end points for sensor events, we leverage the \textit{last sensor fired representation}~\cite{van2008accurate}, which means the last triggered sensor continues to retain its value as 1 and changes to 0 when the next sensor is triggered.
In our data format, the transition between \textit{S$_n^i$} and \textit{S$_n^i+1$} represents this information, that is, the sensor triggered by resident \textit{Pn} is changing. And the corresponding switching timestamp refers to \textit{critical timestamp}.

\subsubsection{Smoothing}
To obtain critical timestamps, we select switching time points when transitions occur. 
However, this processing step includes both real activity switching time and noise values. 
The sensor data is sampled every second. 
The high-density data includes several kinds of unexpected noise values, \eg~ break in continuous activity, loss of data, and so on. 
These noisy values should be smoothed before the final mapping stage.

\subsubsection{Mapping} 
By understanding and learning the relation between activities and sensors, we try to figure out how the system performance will be affected by the sensor layout.

\begin{table}[t]\scriptsize
  \caption{Descriptions of synthetic multi-occupancy models and the comparison with the ARAS dataset}
  \label{tab:multi}
    \begin{tabular}{cccccc}
    \midrule
            & Area Size & Floorplan & \tabincell{c}{Sensor Quantity\\(Sensor Density)} & \tabincell{c}{Sensor Event \\ Occurrences} \\
    \midrule
            ARAS~\cite{alemdar2013aras} & 50m$^2$ & \tabincell{c}{1 bedroom, 1 bathroom, \\ 1 kitchen, 1 living room}  & \tabincell{c}{20 \\ 0.4 sensor/m$^2$} & 187\\
    \midrule
            \tabincell{c}{2-Person \\ Scenario} & 100m$^2$ & \tabincell{c}{2 bedrooms, 1 bathroom, \\ 1 kitchen, 1 living room} & \tabincell{c}{43 \\ 0.43 sensor/m$^2$} & 190 \\
    \midrule
            \tabincell{c}{3-Person \\ Scenario} & 118m$^2$ & \tabincell{c}{3 bedrooms, 1 bathroom, \\ 1 kitchen, 1 living room} & \tabincell{c}{51 \\ 0.43 sensor/m$^2$} & 253 \\
    \midrule
            \tabincell{c}{4-Person \\ Scenario} & 132m$^2$ & \tabincell{c}{4 bedrooms, 1 bathroom, \\ 1 kitchen, 1 living room} & \tabincell{c}{60 \\ 0.45 sensor/m$^2$} & 420 \\
    \midrule
           \tabincell{c}{5-Person \\ Scenario} & 150m$^2$ & \tabincell{c}{5 bedrooms, 1 bathroom, \\ 1 kitchen, 1 living room} & \tabincell{c}{69 \\ 0.46 sensor/m$^2$} & 565 \\
    \midrule
    \end{tabular} \\
\end{table}

\subsection{Synthetic Multi-person Behavior Models}

We leverage five different single-occupancy published datasets~\cite{cook2013transfer} to build synthetic multi-person datasets in this paper.
The occupants from each dataset will act as residents in our emulated environment with their realistic activity data and learned patterns in different multi-person scenarios.

\subsubsection{Properties of the dataset}
Table~\ref{tab:multi} illustrates the properties of several multi-occupancy scenarios and infrastructures' detail in respective emulated environments.
The synthetic multi-person datasets we used for further modeling contains information as described in the section~\ref{subsubsec:format_data}, consisting of important timestamps and sensor ID triggered by residents.
Figure~\ref{fig:sensorlist} shows the sensor activation list in our four-person scenario. 
Four colored lines represent four residents in the space, respectively.

\subsubsection{Validation of the methodology}
\label{subsubsec:valid}
\begin{figure}[t]
     \centering
     \begin{subfigure}[c]{0.23\textwidth}
         \centering
         \includegraphics[width=\textwidth]{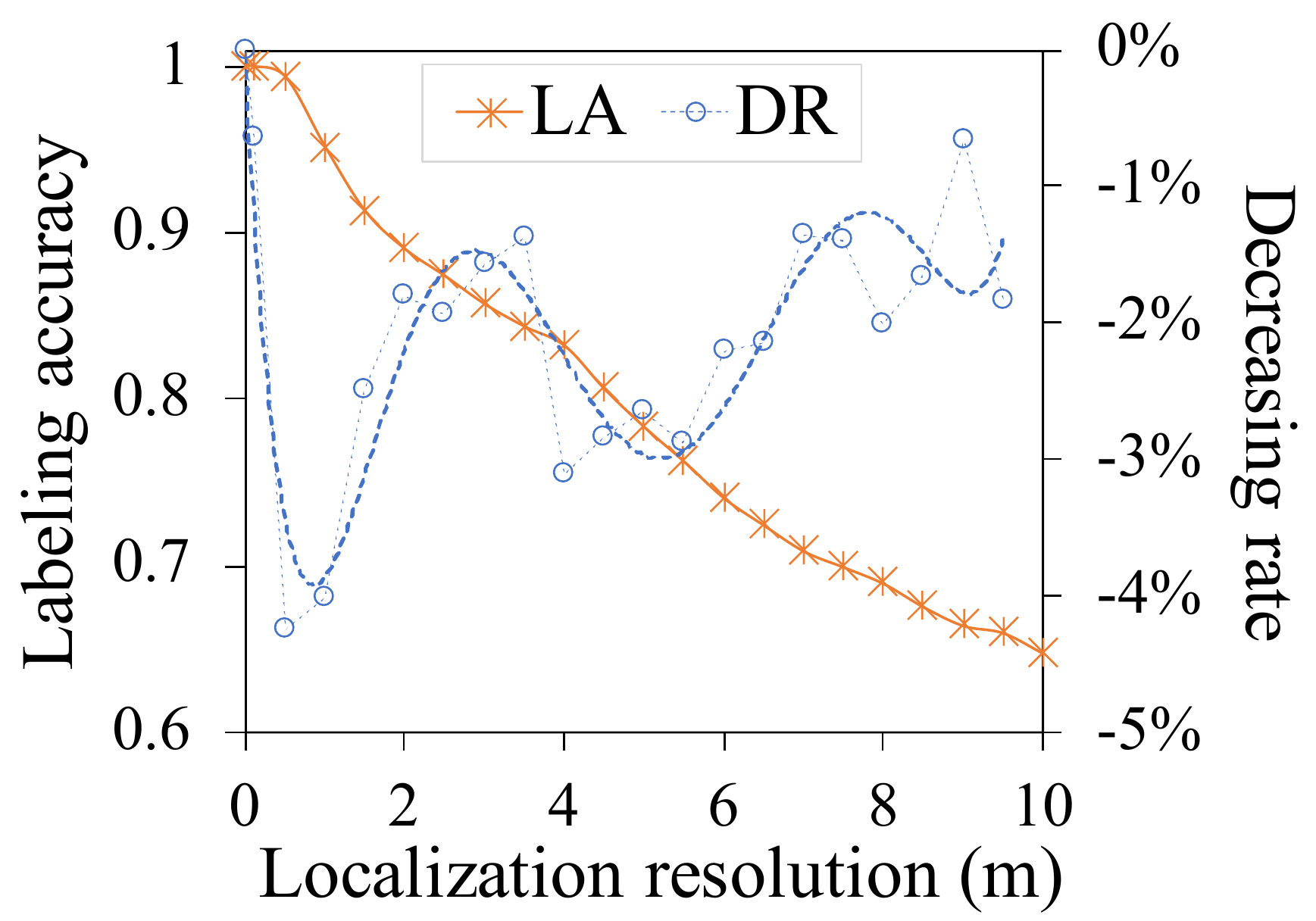}
         \caption{Configuration I}
         \label{fig:I}
     \end{subfigure}
     \begin{subfigure}[c]{0.23\textwidth}
         \centering
         \includegraphics[width=\textwidth]{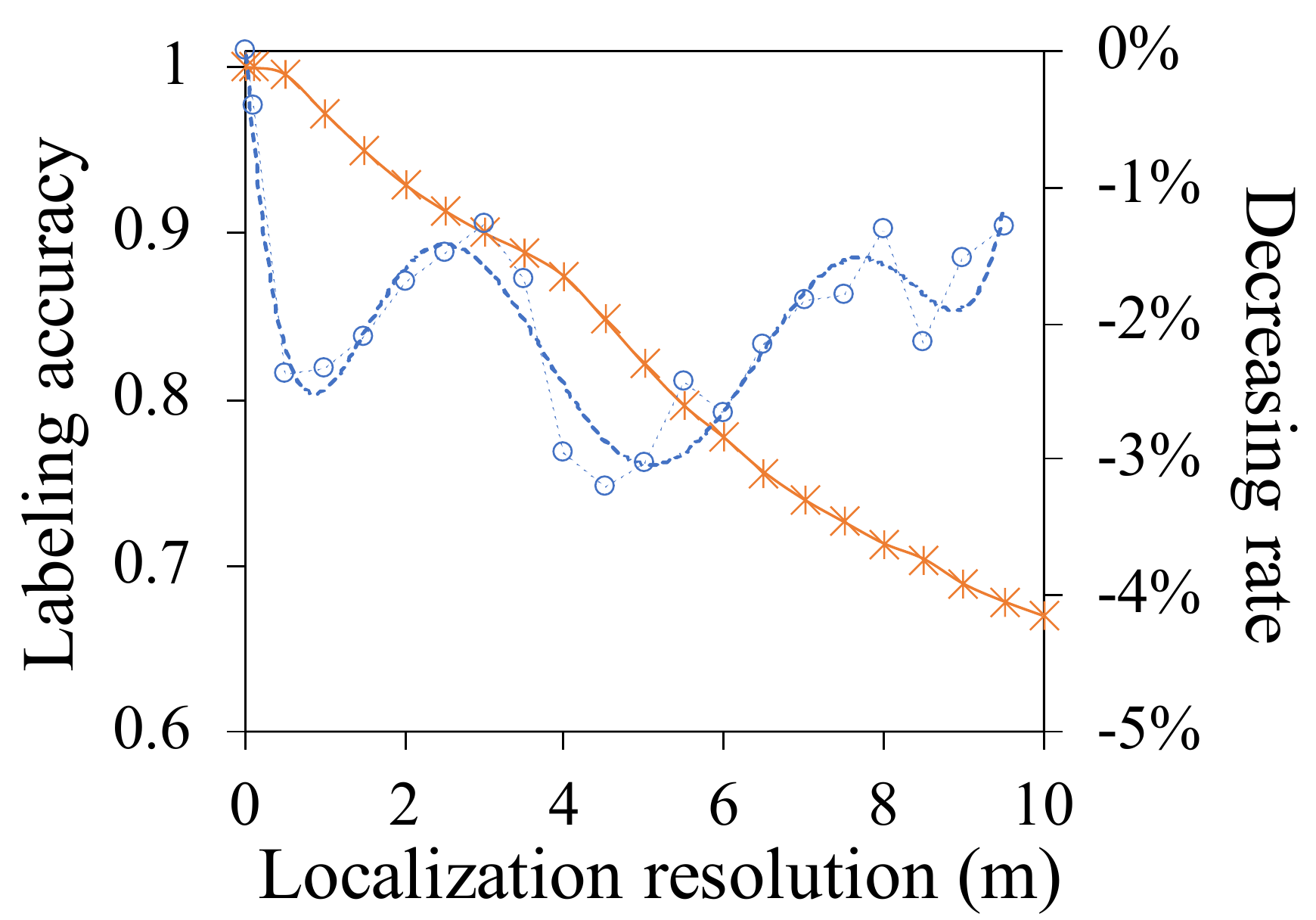}
         \caption{Configuration II}
         \label{fig:II}
     \end{subfigure}
     \begin{subfigure}[c]{0.23\textwidth}
         \centering
         \includegraphics[width=\textwidth]{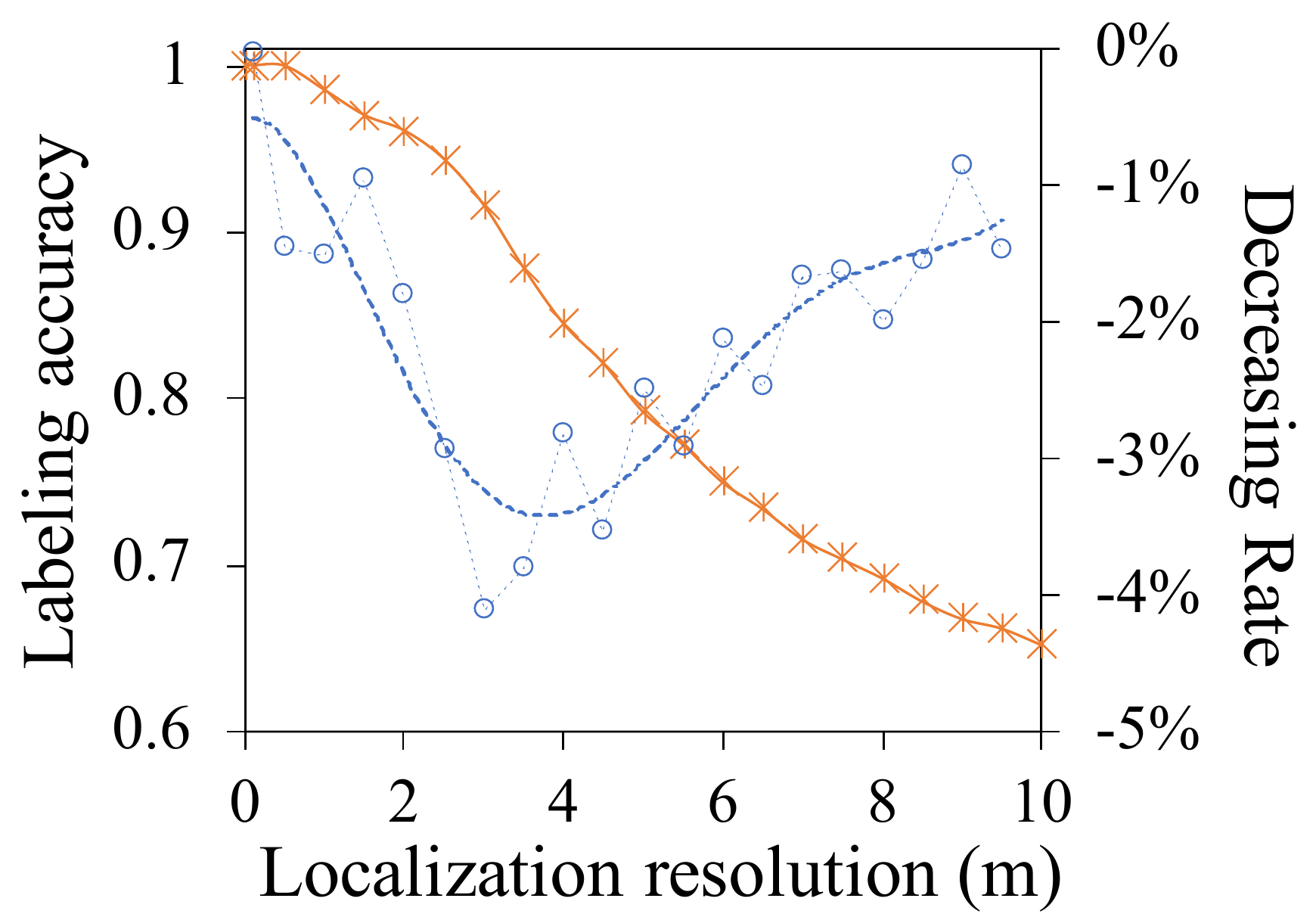}
         \caption{Configuration III}
         \label{fig:III}
     \end{subfigure}
     \begin{subfigure}[c]{0.23\textwidth}
         \centering
         \includegraphics[width=\textwidth]{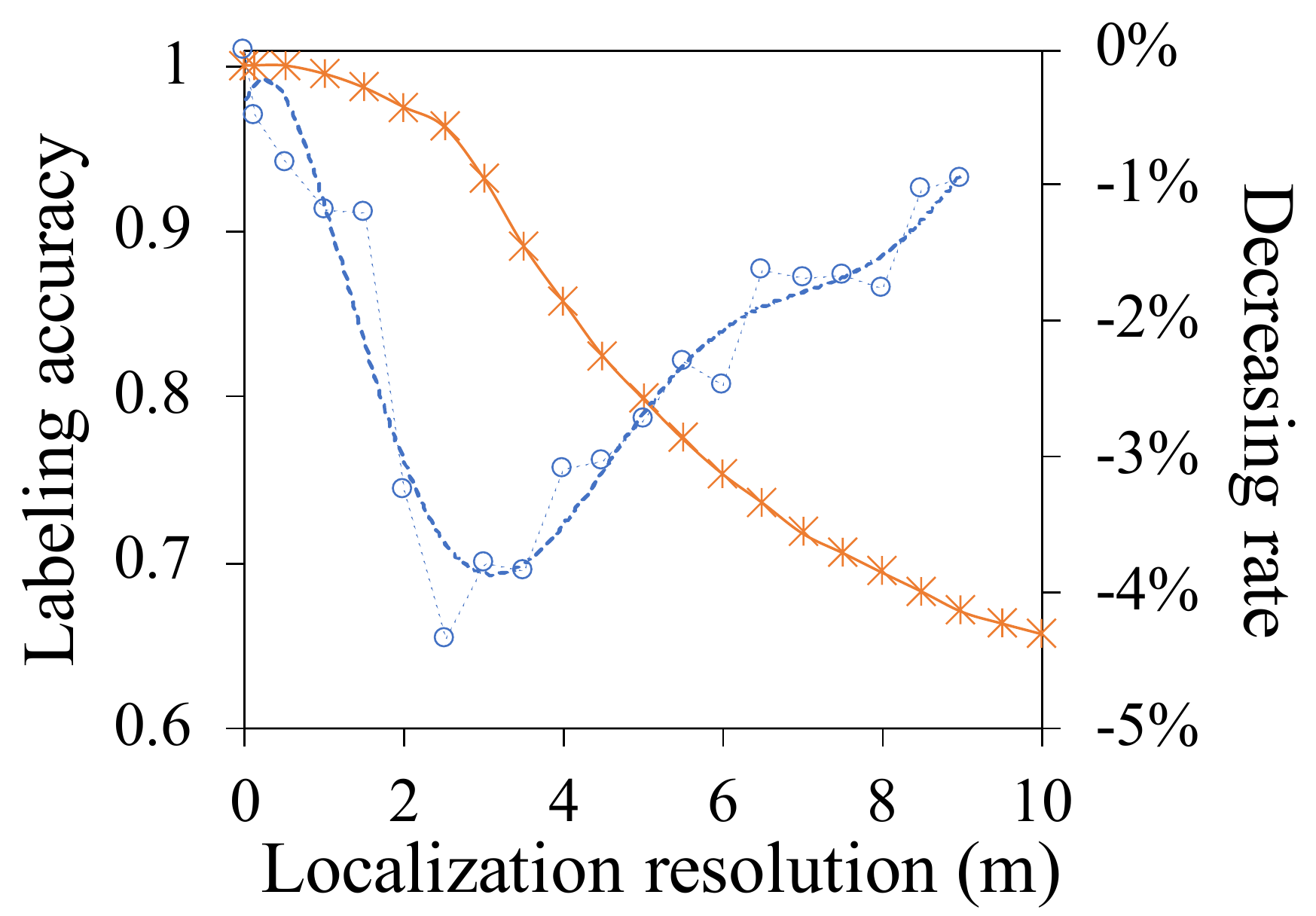}
         \caption{Configuration IV}
         \label{fig:IV}
     \end{subfigure}
        \caption{Similarities between four different configurations
        }
    \label{fig:4Configurations}
\end{figure}

In order to evaluate the quality of the synthetic multiple-occupancy datasets and validate whether the proposed synthetic methodology can represent characteristics of the real-environment datasets, we compare the similarity between the ARAS dataset~\cite{alemdar2013aras}, a real two-person dataset, and the synthetic two-person dataset. 
The ARAS dataset in ARAS(real) floorplan is defined as the baseline, referred to \textit{Configuration I} in Figure~\ref{fig:4Configurations}. Then we compare the synthetic dataset in the ARAS floorplan (\textit{Configuration II}), the ARAS dataset in the emulated floorplan (\textit{Configuration III}), and the synthetic dataset in the emulated floorplan (\textit{Configuration IV}) to the \textit{Configuration I}. 


To quantify the variation of localization resolution in different configurations, both labeling accuracy (\textit{LA}) and corresponding decreasing rate (\textit{DR}) of the same annotation task are analyzed, shown in Figure~\ref{fig:4Configurations}.
We first investigate the similarity when using the ARAS (real) and synthetic datasets in the ARAS (real) floorplan, as shown in Figure~\ref{fig:I} and~\ref{fig:II}. 
Even the synthetic dataset is stochastic and has no relation with the real dataset, it performs as well as the real dataset under an identical layout for the annotation task.
We then compare the performance when the ARAS (real) and synthetic datasets with the same emulated floorplan, as the experiment results shown in Figure~\ref{fig:III} and~\ref{fig:IV}, which also exert similar performance in the same task.
This similarity demonstrates that the synthetic datasets we utilized for further analysis are capable to provide convincing explanations for the annotation task in this paper.

\subsection{Trajectory Generation}
\label{subsec:traj}

Based on the sensor activation versus time, we model \textit{the real trajectory} by choosing optimal paths between adjacent sensor activities. 
The \textit{detected trajectory} is synthesized under the real trajectory and the different localization resolution.

\subsubsection{Real Trajectory Generation}
\label{subsubsec:realtrac}

Leveraging the sensor activation lists with timestamps (as shown in Figure~\ref{fig:sensorlist}), where the target resident's locations can be refereed by the sensor layout, a series of location nodes would be considered to generate the \textit{best route} when a resident move from one sensor to another. The \textit{best route} generated is utilized as \textit{real trajectories} for residents in this paper as people always choose the shortest route when they move from one node to another node. 

\begin{figure}[t!]
  \centering
  \includegraphics[width=1.0\columnwidth]{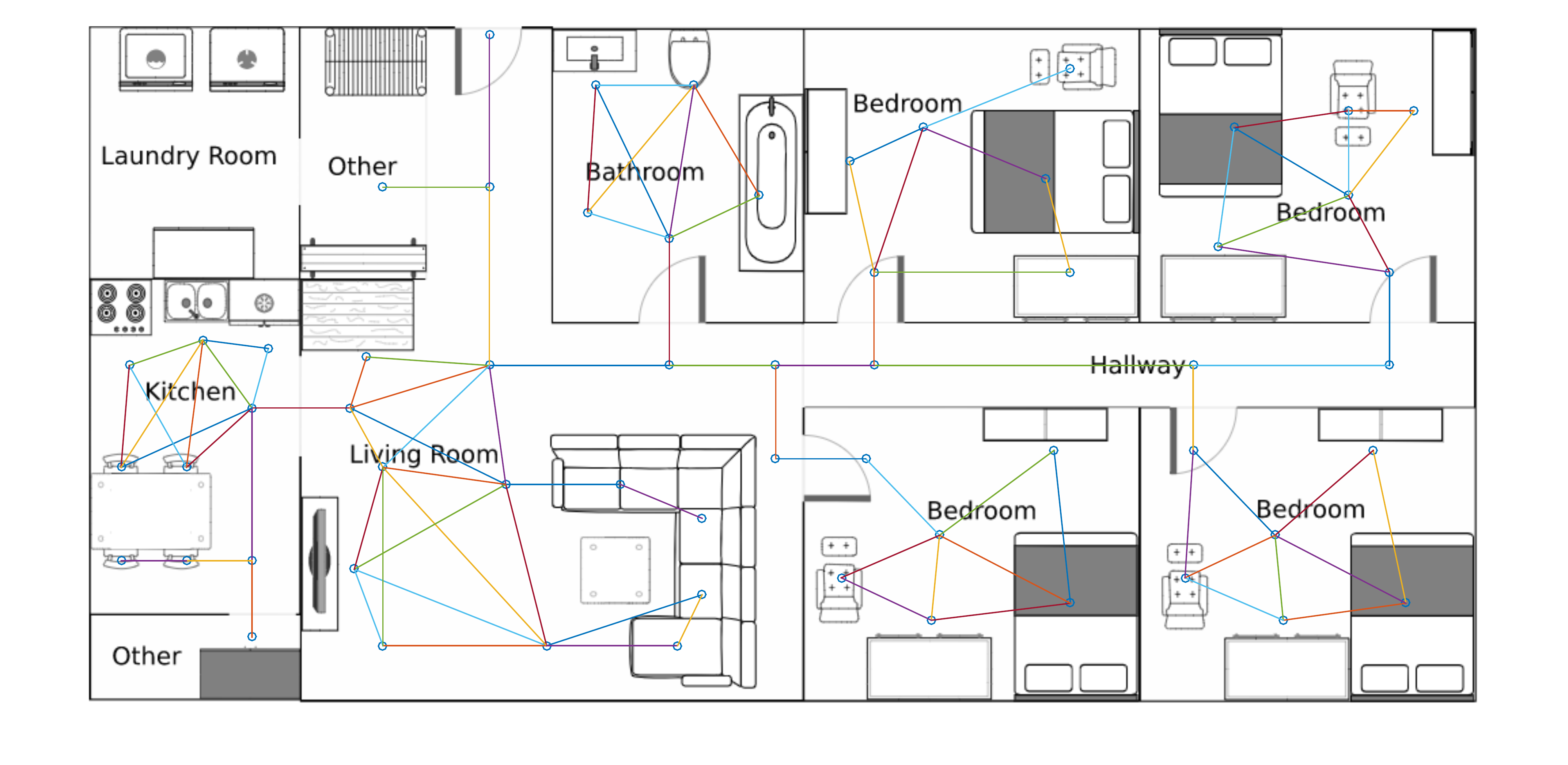}
  \caption{Sensor locations and bridge connections between nodes in a four-person scenario
  }
  \label{fig:bridge}
  \Description{Nodes_Bridge.}
\end{figure}

The nodes include \textit{sensor locations} and \textit{junction locations} to avoid obstacles, as shown in Figure~\ref{fig:bridge}.
Sensors are attached on furniture or appliances that residents are most frequently interact with. 
Some important nodes are added in the \textit{Hallway} and \textit{Living Room} as a transition point from one sensor to another sensor, refer to as \textit{junction locations}. 
\textit{Bridges} represent how people will move from one sensor to the nearby sensors.
If the number of nodes is not big enough, the path will be intercepted by the walls. 
More nodes can result in a smoother moving path but will increase computational cost. 
Given nodes layout and possible path connections (edges), the \textit{bridge map} can be graphed (in Figure~\ref{fig:bridge}).
The \textit{bridge map} includes all possible optimal ways in the room to move from one activity (sensor) to other ones. 
Finding the shortest way between two sensors can be treated as a classic optimization problem. 
To find an optimal path in 2D environments, there are several widely evolved navigation algorithms, e.g. \textit{Dijkstra's algorithms}~\cite{dijkstra1959note}, \textit{A*}~\cite{hart1968formal}, and \textit{Depth First Search (DFS)}~\cite{tarjan1972depth}. 

\begin{figure}[t!]
  \centering
  \includegraphics[width=1.0\columnwidth]{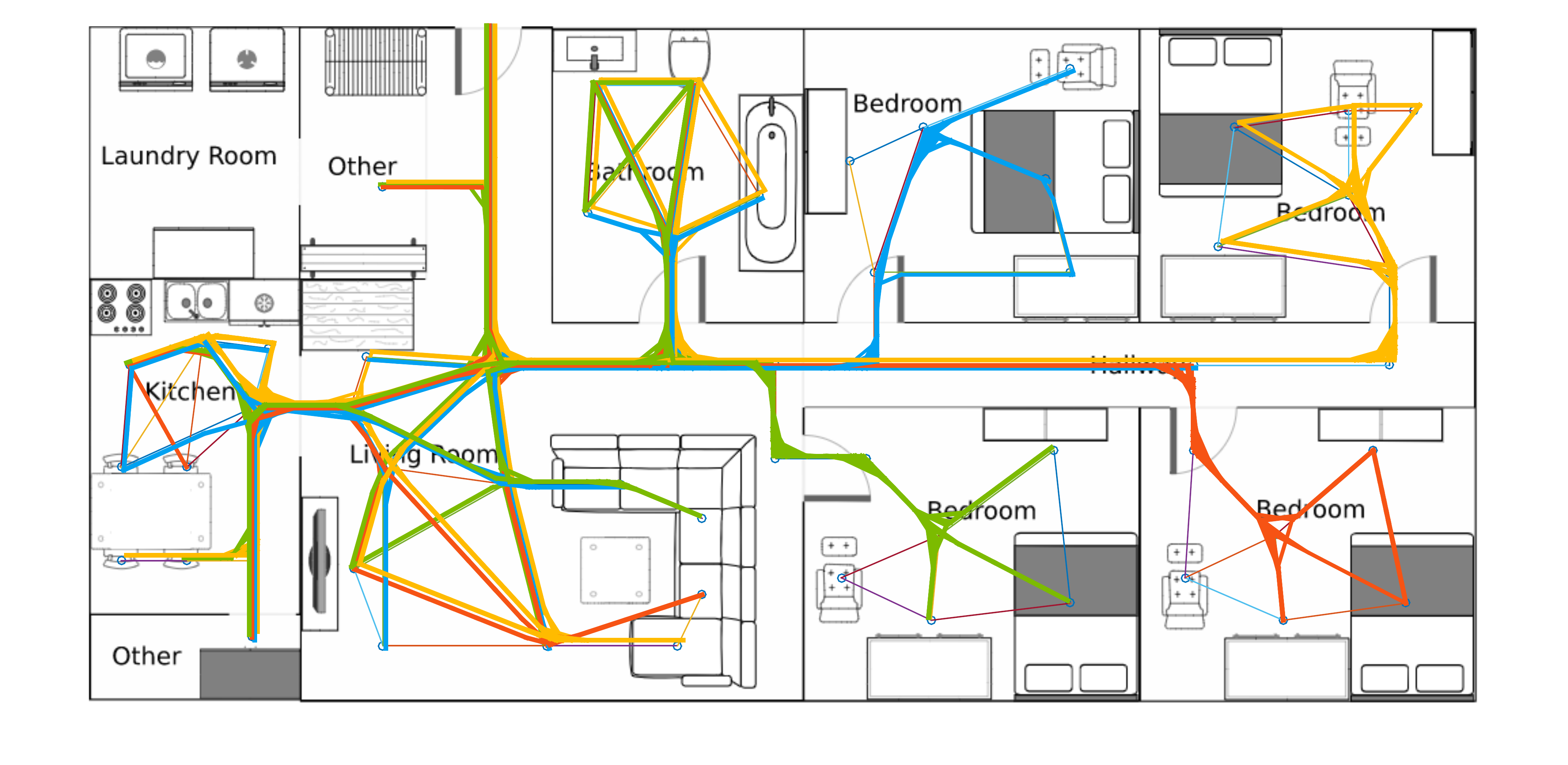}
  \caption{Residents move and interact with the environment in a day using shortest routes; solid lines with red, green, blue and yellow color represent the trajectories of resident A, B, C and D, respectively. 
  }
  \label{fig:realtrac}
  \Description{RealTrac.}
\end{figure}

\textit{Dijkstra's algorithm} is a technique to find the optimal and shortest paths between two different nodes (i.e., starting point and destination point). 
To accelerate the calculation, a \textit{modified Dijkstra's algorithm} is adopted to synthesize the resident's \textit{real trajectory} in this study. Setting the nodes $SL = \{1,\cdots,n\}$, the possible connections $E = \{1,\cdots,m\}$ , the selection of optimal path can be done in $O\left(E + SL \log SL\right)$ step for each selection. 
During the selection, each edge should have one or multiple weights. 
The weights are used to evaluate the capacity and priority of edges. 
In this work, the only considered parameter is the distance, without additional priority factors. 
Each path can be endowed priorities depending on the specific case to improve further.
Figure~\ref{fig:realtrac} shows the emulated \textit{real trajectories} based on four residents' sensor activation lists, respectively.

\subsubsection{Detected Trajectory Generation}
The RTLS devices can provide the \textit{detected trajectory} within the respective resolution\cite{brena2017evolution,aguilera2018broadband,nadeem2014highly,xu2017rfid,pan2015indoor}. 
The localization range and resolution of the facilities depend on the different indoor positioning technologies~\cite{ruiz2017comparing}.
After the \textit{real trajectory} is generated, the possible resolution is incorporated in the trajectory with a given device (sensor) performance. 
Assuming the localization resolution is $L$ and the possibility of each point in the error range is identical, the error can be added as Equation \ref{error}.

\begin{linenomath*}
\begin{equation}
\left\{
    \begin{array}{lr}
    x = x_r + E\cdot L\cos \theta_i &  \\
    y = y_r + E\cdot L\sin \theta_i
    \end{array}
\right.
\label{error}
\end{equation}
\end{linenomath*}

\noindent where $\theta_i$ is the possible angle for sensor $i$ from real location $\left(x_r,y_r\right)$, $\left(x,y\right)$ is the detected location and $E\in\left[0,1\right]$ is error factor.

In this study, the distribution of added error is assumed uniform in all locations of the shooting range. 
Error factor $E$ can be any value from zero to one with identical possibility. 
If the resolution of a particular device (sensor) has a specific pattern (\eg~ \textit{semi-normal distribution}), the error factor $E\left(x\right)$ should be controlled to have a changing possibility along with the distance from the sensor. 
If the sensor is attached on a wall, resulting in a sector detected range, the detecting location is controlled in a corresponding direction. 

\begin{figure*}[t!]
     \centering
     \begin{subfigure}[b]{0.33\textwidth}
         \centering
         \includegraphics[width=1.1\textwidth]{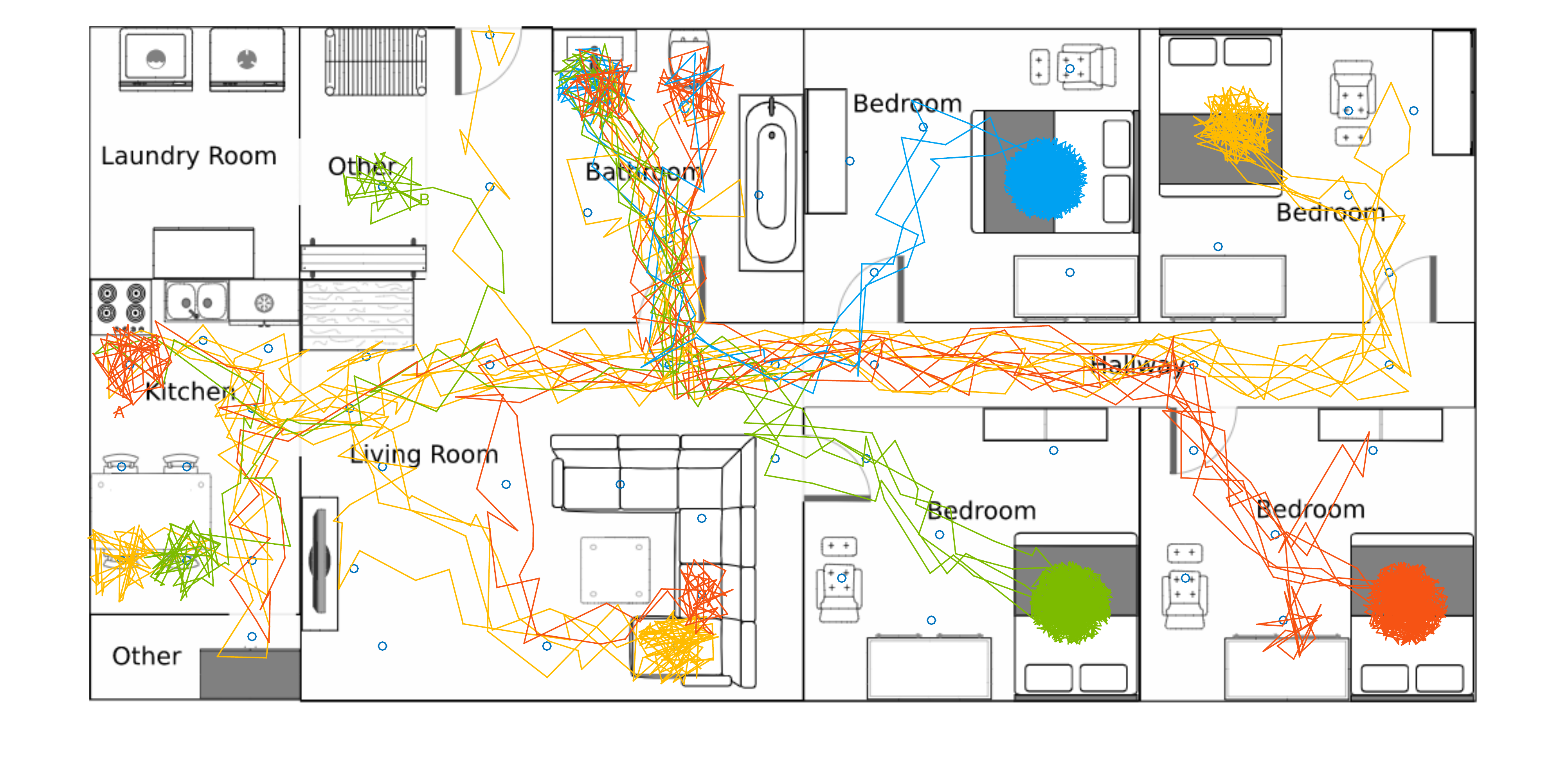}
         \caption{Time 00:00-08:00}
         \label{fig:time1}
     \end{subfigure}
     \begin{subfigure}[b]{0.33\textwidth}
         \centering
         \includegraphics[width=1.1\textwidth]{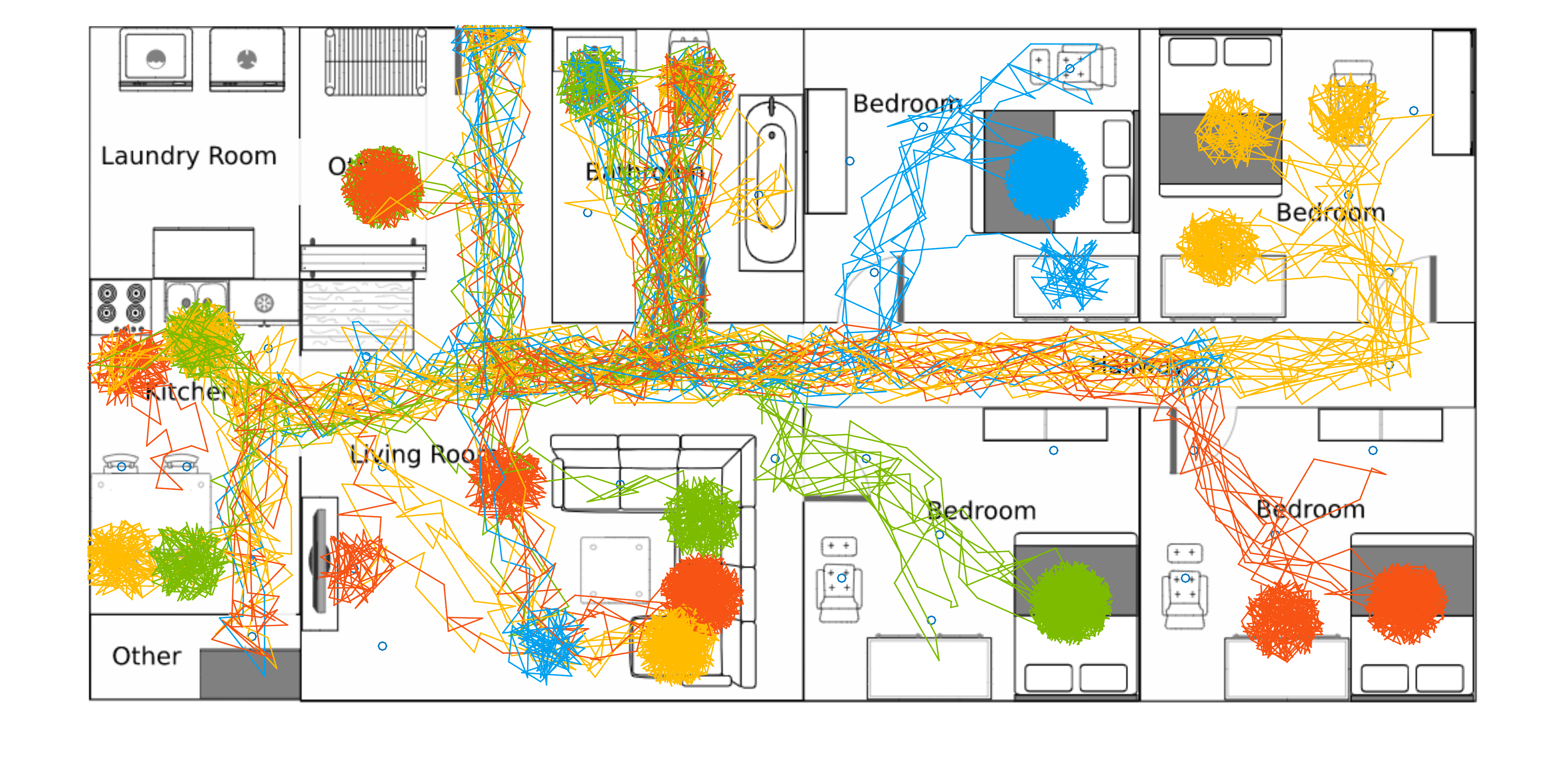}
         \caption{Time 00:00-16:00}
         \label{fig:time2}
     \end{subfigure}
     \begin{subfigure}[b]{0.33\textwidth}
         \centering
         \includegraphics[width=1.1\textwidth]{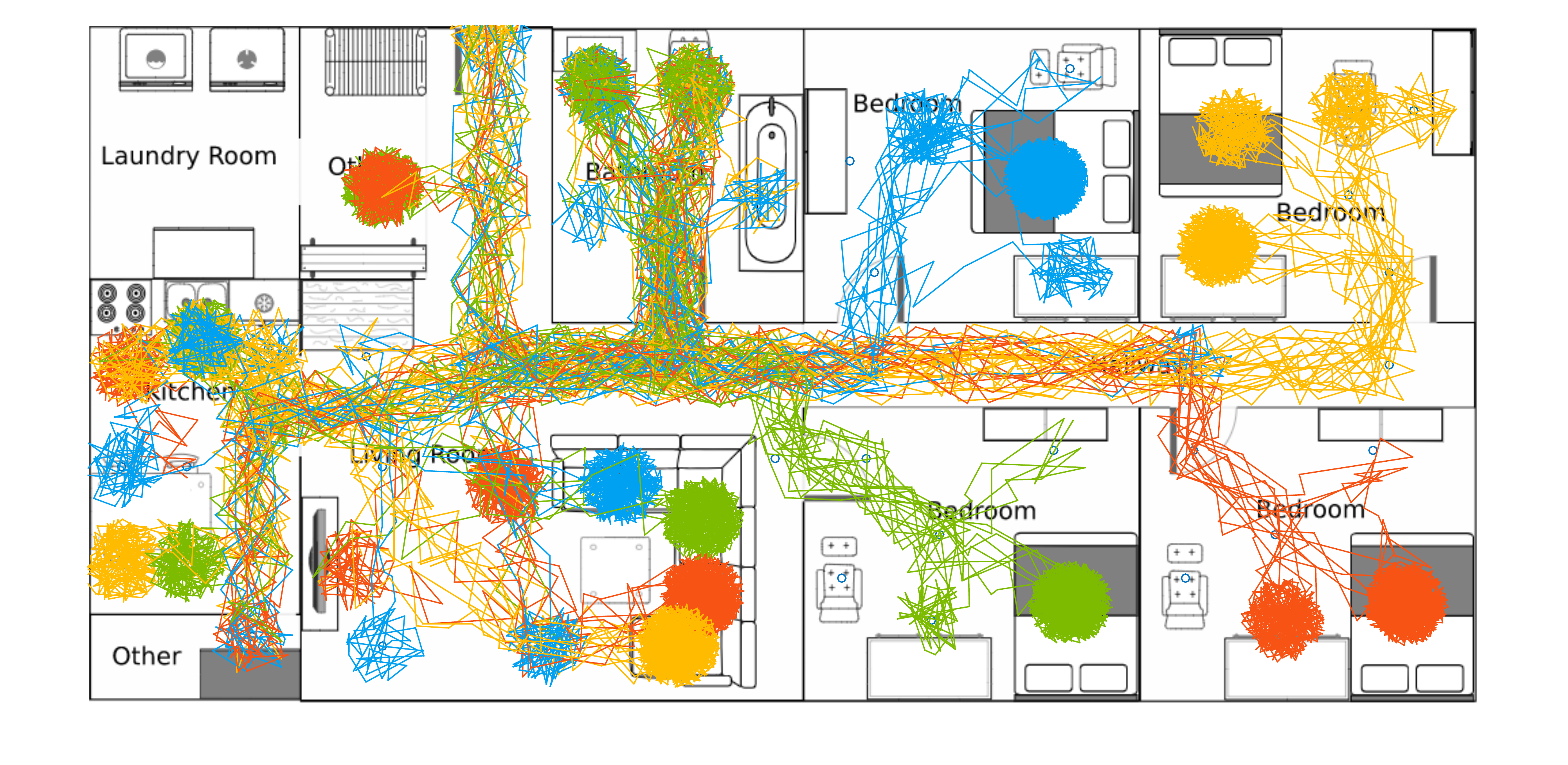}
         \caption{Time 00:00-23:59}
         \label{fig:time3}
     \end{subfigure}
     \caption{Emulated deviations (0.5-meter resolution) are added in the route for four residents. Solid lines with red, green, blue and yellow color represent the trajectories of resident A, B, C and D, respectively. 
     }
        \label{fig:threetime} 
\end{figure*}

Figure~\ref{fig:threetime} shows a representative trajectory \textit{(L = 0.5 meters)}, that is how \mosen~emulates localization devices collect resident's data in a day. 
It gives a straightforward insight into how residents have different activity patterns that live in a single environment, which can be quite intricate.

\section{Evaluation}
\label{sec:evluation}

\subsection{Mosen Implementation}
Typical studies in multi-occupancy scenarios may consider two residents in the same space, limited by costly devices, annotation problem, localization accuracy, or others. 
With our \mosen~platform, we investigate different multi-person scenarios by emulating independent residents in \textit{any} target sensor network.
We then provide the sensor selection strategy for sensor network by balancing the trade-off between deployment cost and system performance.
The results for our experiments would inform real sensor deployment of the multi-resident smart home.
\subsubsection{Design strategy of multi-occupancy environment}
As the scarcity of multi-occupancy activity dataset in the sensor-based setting, there are few research studies considering more than two residents in a single space.
Leveraging two real two-person datasets, noted as ARAS~\cite{alemdar2013aras} and CASAS~\cite{cook2013transfer}, we validate the efficiency of our two-person data generator in Section~\ref{subsubsec:valid}.
We expand the generator and emulate multi-occupant scenarios in households with 2 to 5 residents.

\mosen~platform is adaptable to different or customized indoor environments, and the critical inputs to the platform are the sensor locations in the target floorplan.
In this evaluation, one design strategy chosen for the increasing multi-person scenarios is to maintain \emph{similar} layout complexity and sensor density, only the number of bedrooms changes with the number of residents.
Figure~\ref{fig:sensorlayout} shows the representative floorplan in the four-person scenario and Table~\ref{tab:realsyn} demonstrates the space size of each scenario.
Generally, every multi-person environment in this evaluation consists of:
\\\textit{Living room, kitchen, laundry room, bathroom, hallway, bedrooms (corresponding quantity is depending on the number of residents) and others.}

\subsubsection{Integration}
Synthetic multi-person datasets and multi-occupancy environments are integrated to emulate residents' indoor trajectories. 
The emulation idea is based on when people move between two triggered sensors, they probably would choose the shortest way from the current sensor location to the next. This shortest way is referred to as the \textit{best route} in this paper, as discussed in the Section~\ref{subsubsec:realtrac}.
We leverage these best routes as residents' \textit{real trajectories}, while the \textit{detected trajectories} are emulating how different positioning resolutions of sensor network track human activities.

However, different positioning technologies with respective resolutions exert varying accuracy when these technologies locate residents in a real environment, which also has diverse deployment costs~\cite{mendoza2019meta}.
For instance, when detecting human locations and trajectories, there are deviations from the ground-truth values, caused by different resolutions. 
For example, Estimote~\cite{mycek2017systems} announced their location beacons could achieve 1.5 meters accuracy, which means that the detected location and their real location might be away from each other at most 1.5 meters.
In this paper, with varying resolutions of different technologies (ranging from 0 meter to 10 meters), we add respective deviations to the \textit{best route}, refer the new trajectories as the \textit{detected trajectories} that are obtained by the distinct localization devices. 

\begin{table}[t]\footnotesize
  \caption{The percentage of labeling accuracy in different scenarios by increasing the number of residents, with different localization resolutions, from 0.5-meter to 10-meter localization resolution.
  }
  \label{tab:overall}
    \begin{tabular}{ccccc}
    \toprule
            Resolution (meters) & 2-Person & 3-Person & 4-Person & 5-Person \\
    \midrule
            0.5 & 100\% & 100\% & 99.89\% & 99.64\% \\
    \midrule
            1 & 99.64\% & 99.62\% & 99.35\% & 98.68\% \\
    \midrule
            2 & 98.19\% & 98.39\% & \textbf{94.56\%} & \textbf{94.03\%} \\
    \midrule
            3 & 95.57\% & \textbf{93.88\%} & \textbf{88.16\%} & \textbf{86.71\%} \\
    \midrule
            4 & \textbf{91.02\%} & \textbf{87.61\%} & 80.49\% & 77.61\% \\
    \midrule
            5 & \textbf{87.46\%} & 82.52\% & 74.57\% & 70.86\% \\
    \midrule
            6 & 84.64\% & 78.66\% & 70.00\% & 65.83\% \\
    \midrule
            7 & 82.63\% & 75.58\% & 66.64\% & 61.74\% \\
    \midrule
            8 & 81.01\% & 73.33\% & 63.65\% & 58.07\% \\
    \midrule
            9 & 79.52\% & 71.42\% & 60.75\% & 54.99\% \\
    \midrule
            10 & 78.41\% & 69.95\% & 58.13\% & 52.12\% \\
    \bottomrule
    \end{tabular}
\end{table}
\subsubsection{Automatic identification labeling}
Providing personalized services to different residents is one of the most important applications in multi-person smart homes. 
By profiling the resident's activity pattern, recording what sensors they have interacted with in their daily routines are the preliminary work in the multi-person activity recognition. 


Automatic identification labeling is the central problem when we try to model multi-occupancy activity recognition based on ambient-sensor networks.
One feasible solution is to use residents' trajectories to label the triggered sensors by matching the location of the sensor and the resident.
The \textit{Graph and Rule Based Algorithm (GR/ED)}~\cite{crandall2013tracking} and the \textit{nearest neighbor standard filter (NNSF)}~\cite{bar1990tracking} are leveraged in this paper to solve the annotation problem, as details in Section~\ref{sec:overview}.
We compare the detected locations of all residents and the triggered sensor at every critical timestamp.
The triggered sensor, hence, would be assigned to the resident who has the shortest straight-line distance.
As illustrated in Section~\ref{subsec:traj}, in the \mosen~emulation, the \textit{best route} represents the ground-truth locations of every resident, while the \textit{detected trajectories} are utilized in the realistic annotation process.


\begin{figure}[t]
  \centering
  \includegraphics[width=2.5in]{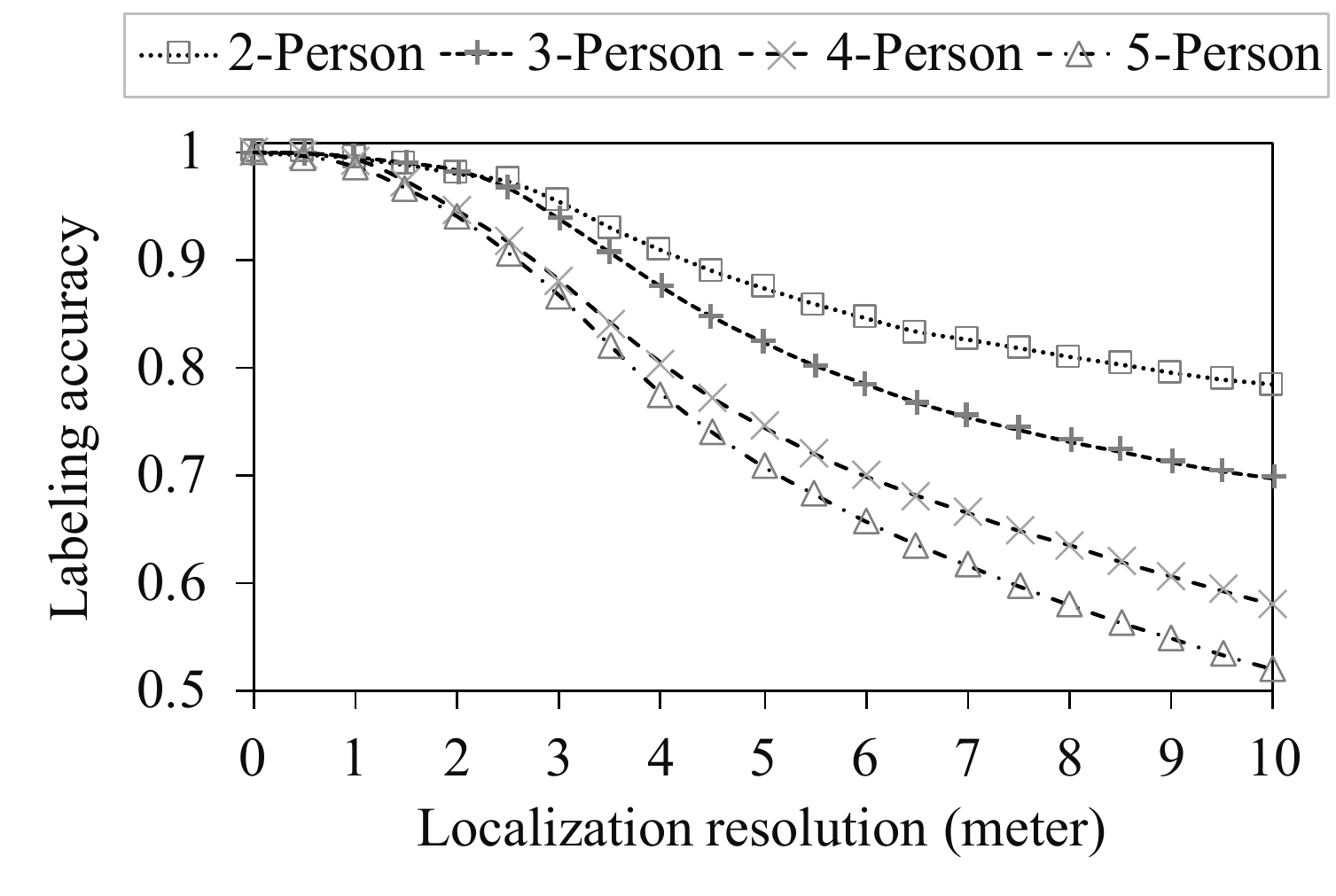}
  \caption{The effect of different localization resolutions on the automatic labeling accuracy in the 2-person, 3-person, 4-person and 5-person scenario, respectively. 
  }
  \label{fig:eloc}
  \Description{eloc.}
\end{figure}

\begin{table}[t]\footnotesize
  \caption{Mean distance and variance for the sensor nodes distribution in several multi-occupancy scenarios
  }
  \label{tab:sensordistance}
    \begin{tabular}{ccccc}
    \toprule
            Scenario & \tabincell{c}{Mean Distance (meter)} & \tabincell{c}{Variance} & \tabincell{c}{Transition Point Interval$^*$} \\
    \midrule
           2-Person & 1.6865 & 0.5305 & 3.0-3.1 meters \\
    \midrule
           3-Person & 1.7065 & 0.5278 & 3.4-3.5 meters \\
    \midrule
           4-Person & 1.6692 & 0.4337 & 3.0-3.1 meters \\
    \midrule
           5-Person & 1.6936 & 0.5714 & 3.2-3.3 meters \\
    \bottomrule
    \end{tabular}
        \scriptsize{\tabincell{c}{\emph{$^*$Transition point interval} refers to the interval that contains the highest decreasing point.}}
\end{table}

\subsection{Performance}

In this evaluation, we choose the automatic identification annotation problem as the central task to illustrate how \mosen~platform evaluate the impact of the number of residents, the number of sensors, positioning technologies, and different sensor layouts. 
\mosen~can additionally provide a \emph{sensor selection strategy} that fits for the user's requirement while optimizing for the number of sensors and their placement (hence the installation cost) to achieve the highest labeling accuracy.

While the real datasets leveraged to model the synthetic multi-occupancy datasets have data for more than one month (Table~\ref{tab:fivebed}), only one-day data from the synthetic datasets are utilized in further analysis. 
We leave an extensive evaluation for weekly data or monthly data for future work.

\begin{figure*}[t]
     \centering
     \begin{subfigure}[c]{0.42\textwidth}
         \centering
         \includegraphics[width=\textwidth]{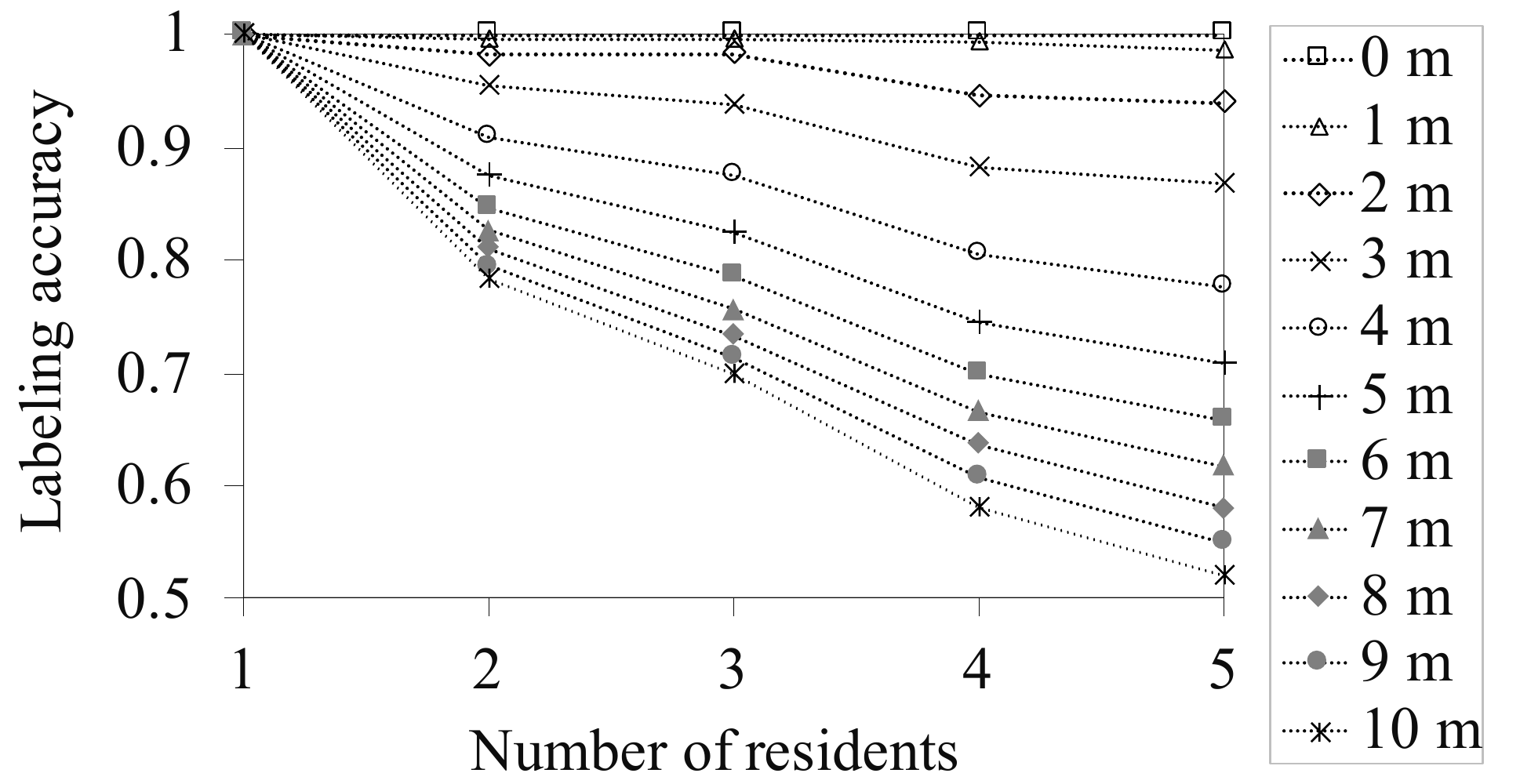}
         \caption{Effect of the residents' quantity}
         \label{fig:overall}
     \end{subfigure}
     \begin{subfigure}[c]{0.28\textwidth}
         \centering
         \includegraphics[width=\textwidth]{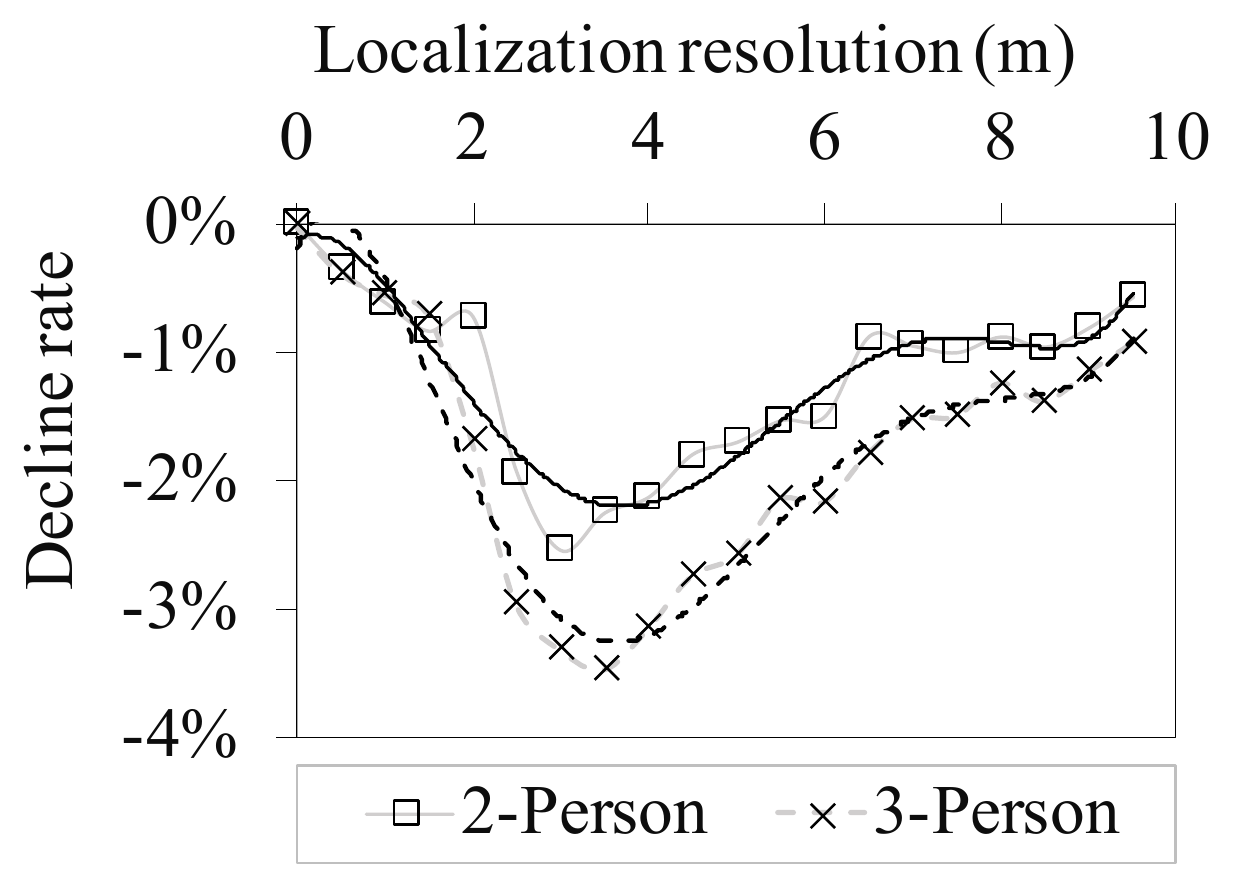}
         \caption{Decline rate with 2 and 3 persons}
         \label{fig:2&3}
     \end{subfigure}
     \begin{subfigure}[c]{0.28\textwidth}
         \centering
         \includegraphics[width=\textwidth]{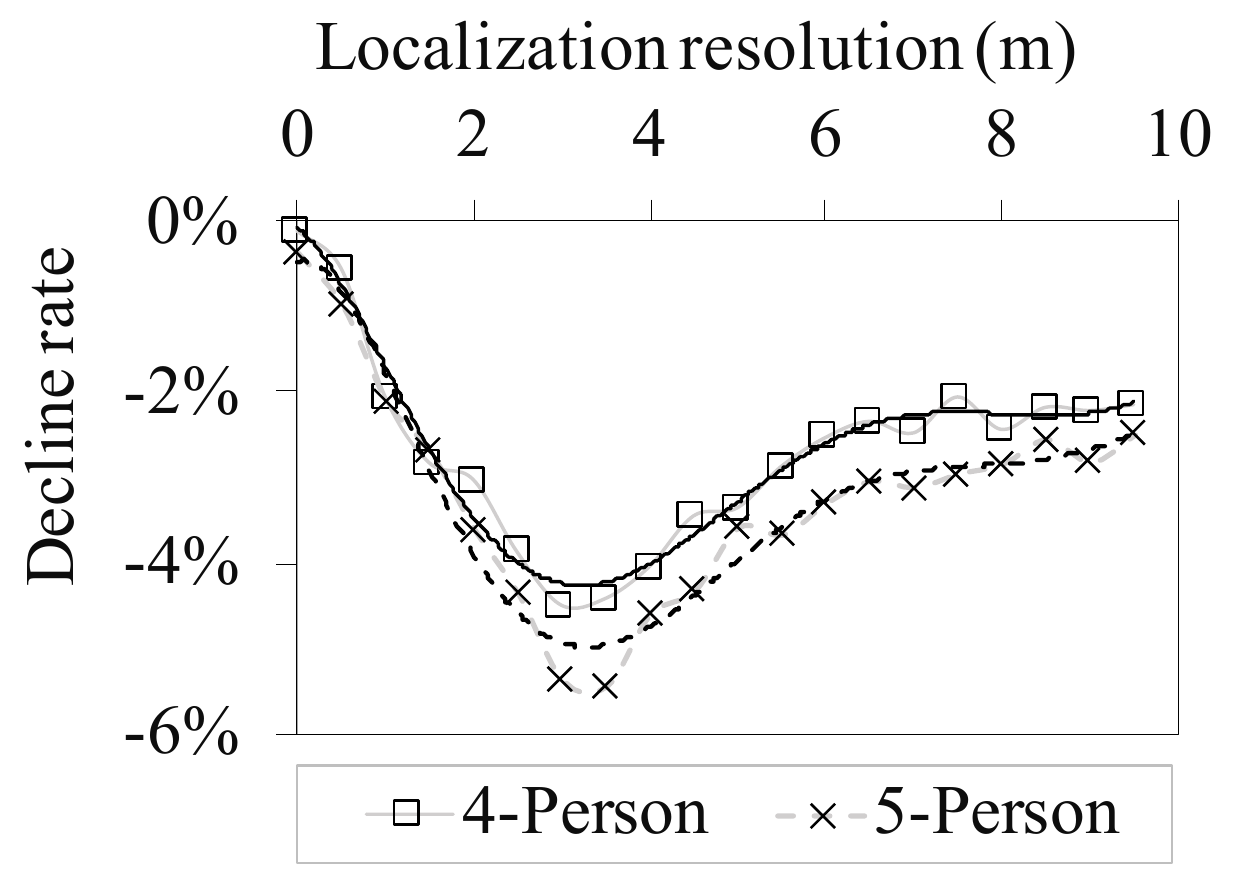}
         \caption{Decline rate with 4 and 5 persons}
         \label{fig:4&5}
     \end{subfigure}
        \caption{Effect of the number of residents on accuracy and decline rate of automatic labeling in different scenarios.}
        \label{fig:effperson}
\end{figure*}

\begin{figure*}[t]
     \centering
     \begin{subfigure}[c]{0.24\textwidth}
         \centering
         \includegraphics[width=\textwidth]{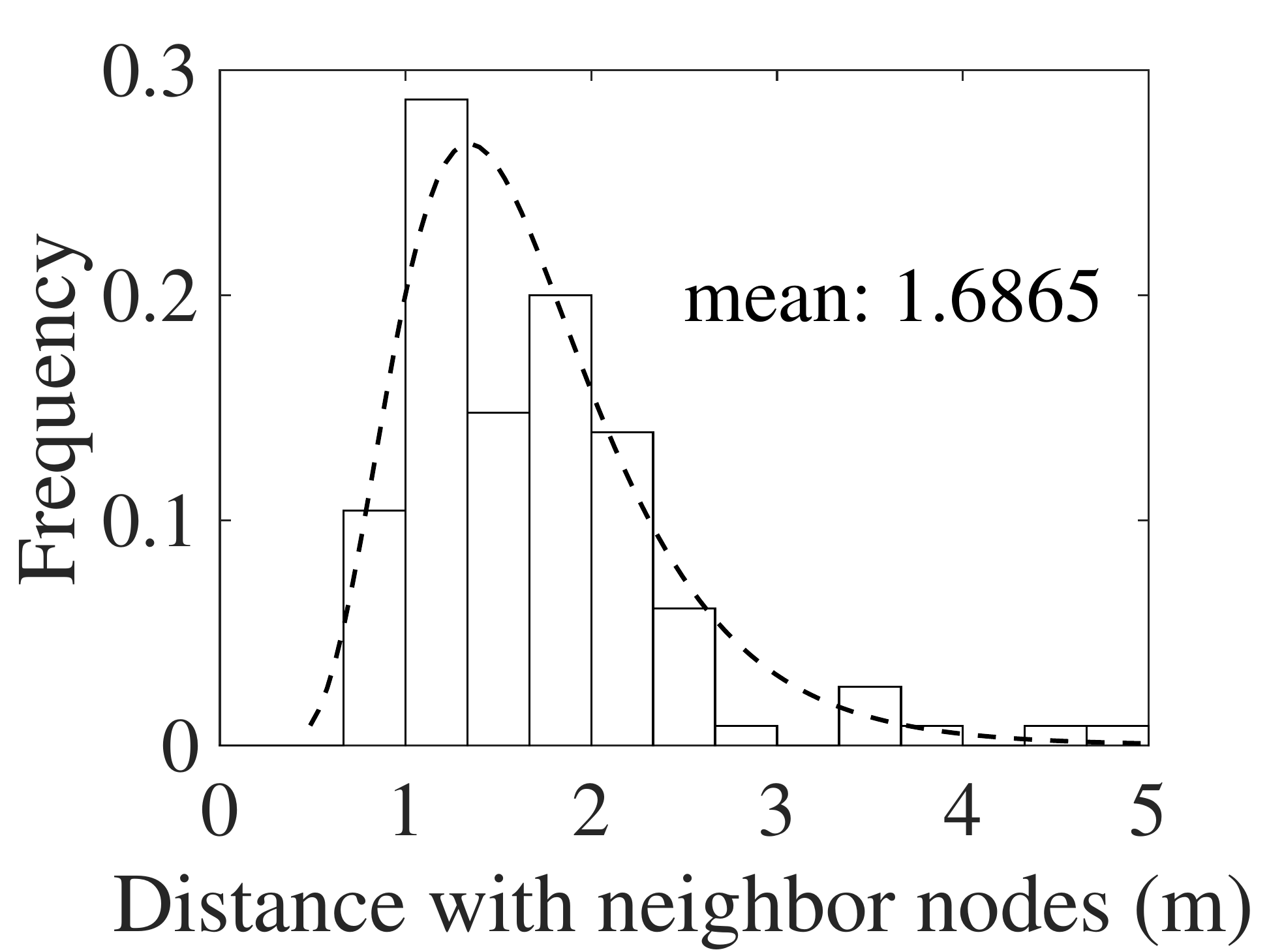}
         \caption{2-Occupancy floorplan}
         \label{fig:2p_dis}
     \end{subfigure}
    \hfill
     \begin{subfigure}[c]{0.24\textwidth}
         \centering
         \includegraphics[width=\textwidth]{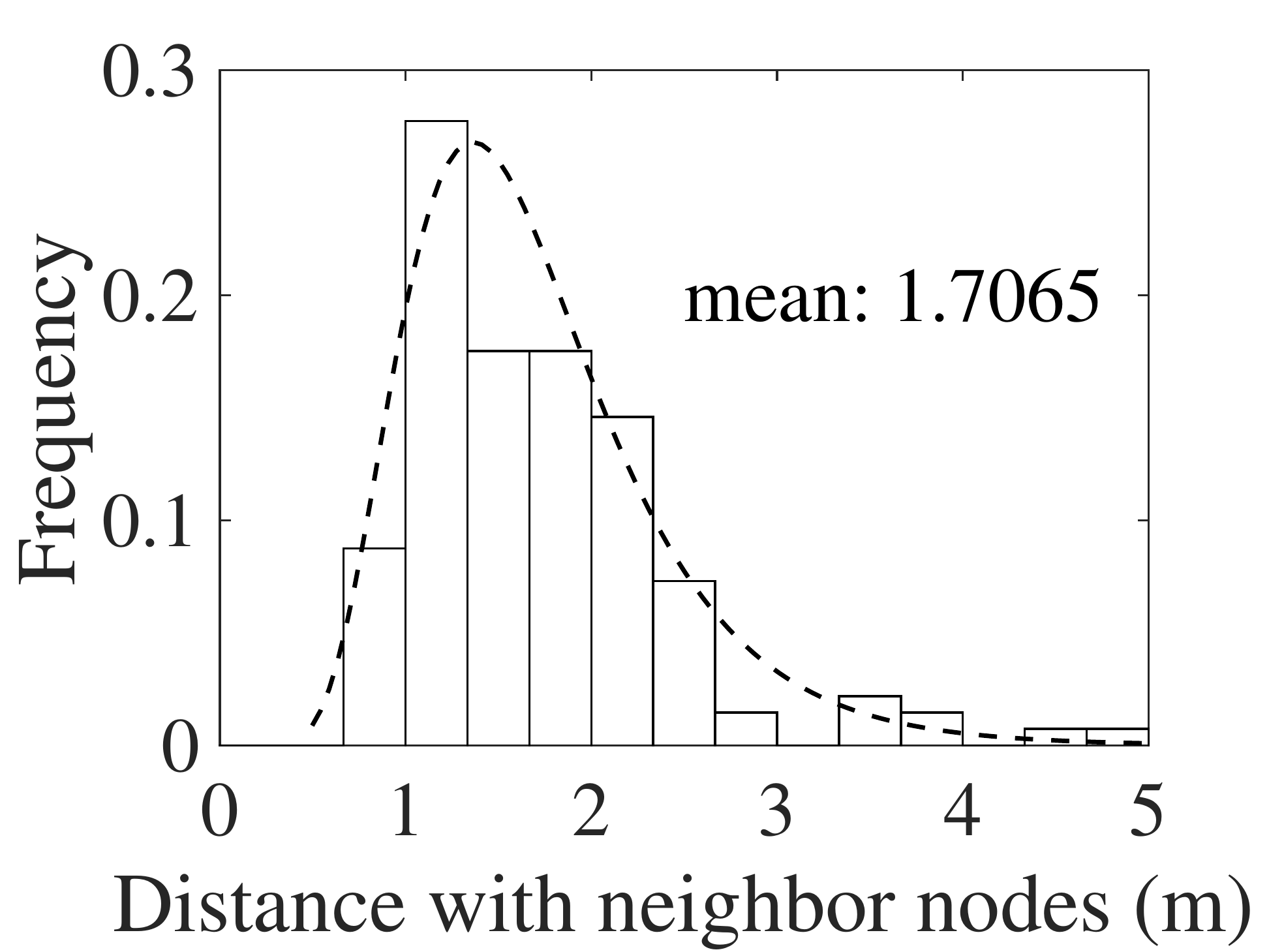}
         \caption{3-Occupancy floorplan}
         \label{fig:3p_dis}
     \end{subfigure}
     \hfill
     \begin{subfigure}[c]{0.24\textwidth}
         \centering
         \includegraphics[width=\textwidth]{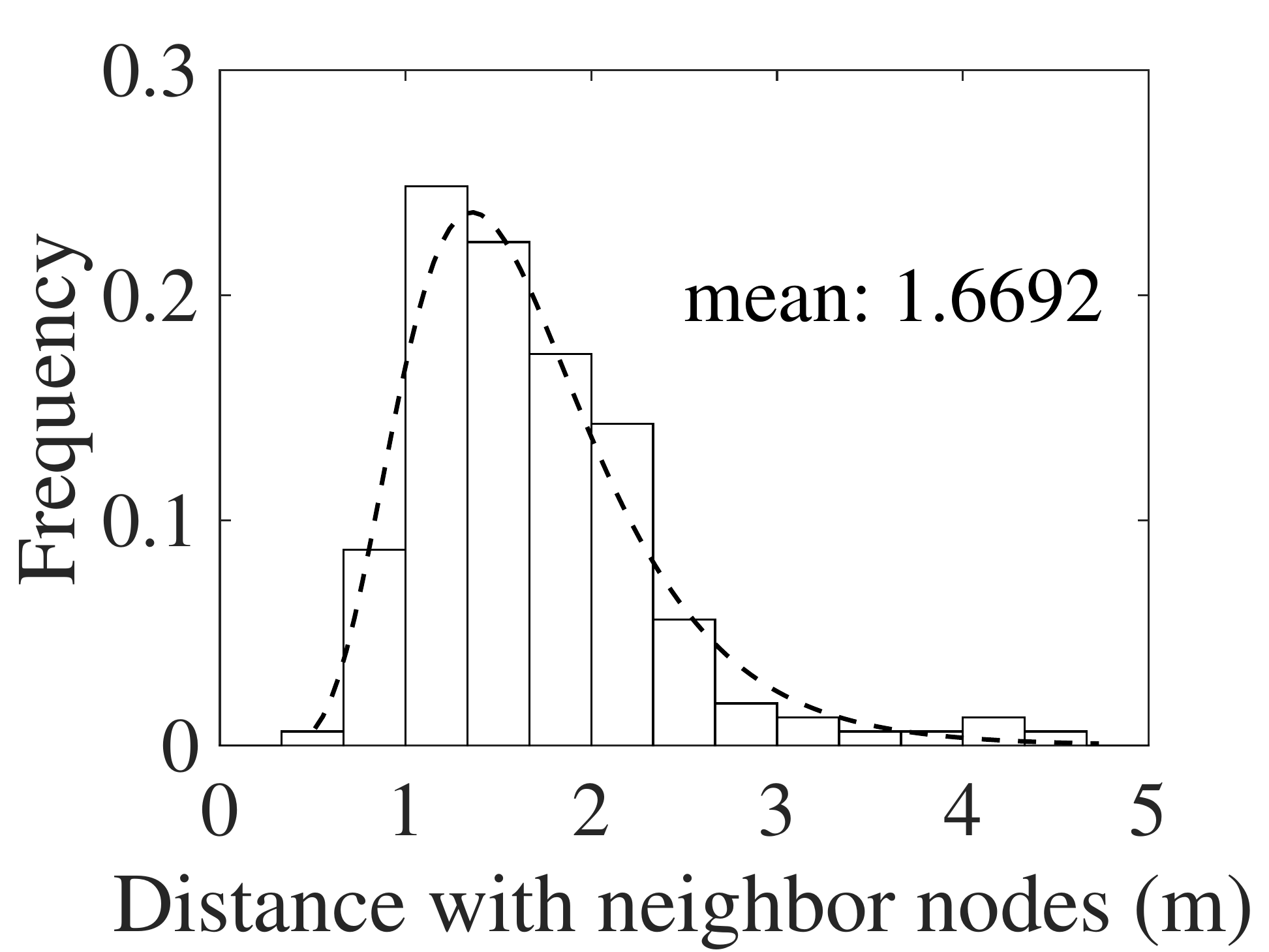}
         \caption{4-Occupancy floorplan}
         \label{fig:4p_dis}
     \end{subfigure}
     \hfill
     \begin{subfigure}[c]{0.24\textwidth}
         \centering
         \includegraphics[width=\textwidth]{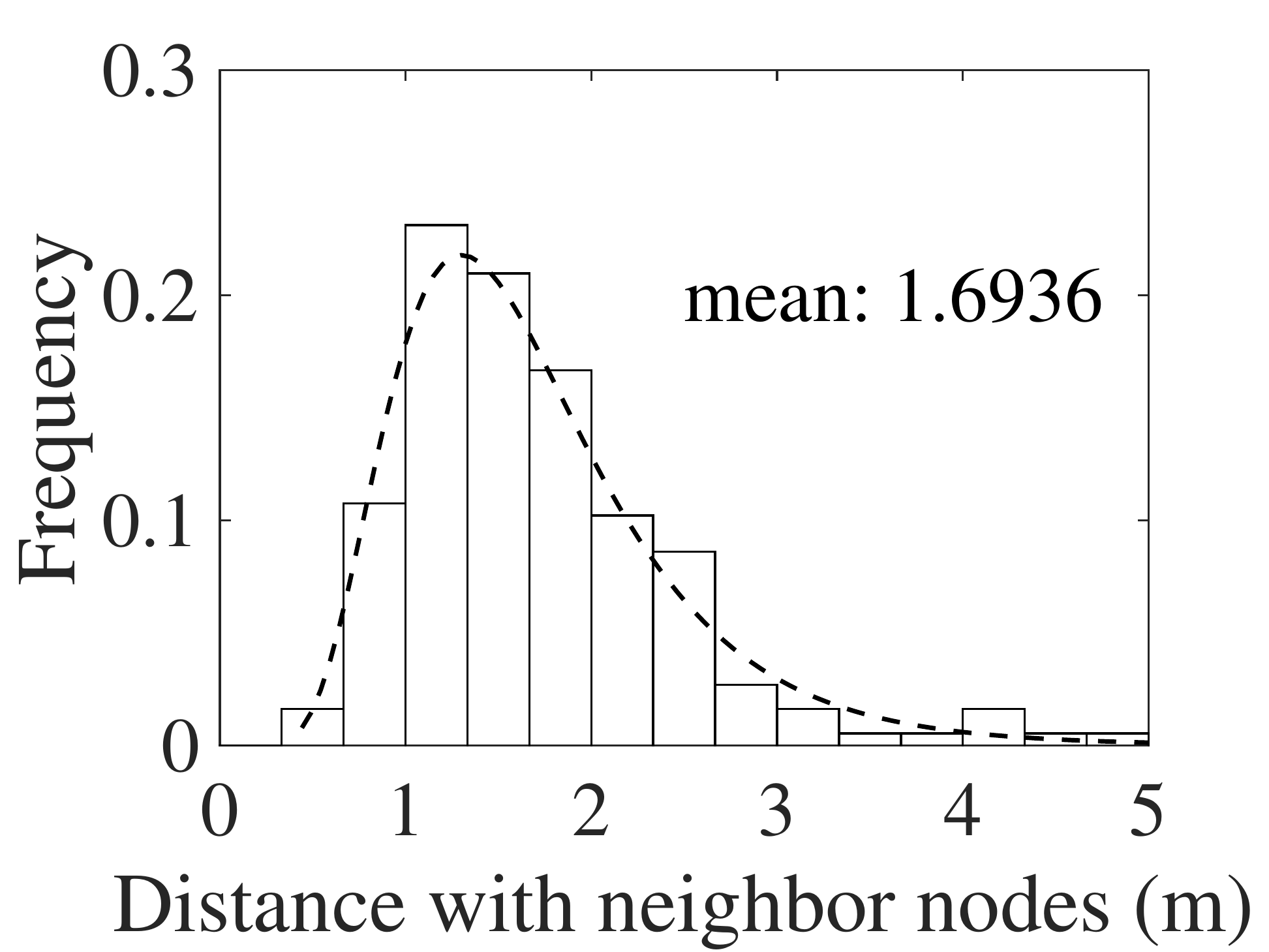}
         \caption{5-Occupancy floorplan}
         \label{fig:5p_dis}
     \end{subfigure}
        \caption{Distributions of the sensor connection length in four multi-occupancy scenarios. The x-axis represents the distance between nodes (sensors), and the y-axis represents the frequency of the corresponding distance between nodes.}
        \label{fig:sensordistribution}
\end{figure*}

\subsubsection{Labeling Performance}

In Table~\ref{tab:overall}, we first compare the performance of automatic identification labeling in several multi-occupancy scenarios.
We use \emph{similar} floorplans and sensor layouts to emulate multiple residents living in a single space. 
It generally provides information about the selection of localization devices when an application has a different demand on labeling accuracy. 
The labeling accuracy can be improved with a smaller resolution, but the cost of devices will also increase. 
Insights from this table could provide valuable information to designers or practitioners when designing the actual sensor deployments. 
For example, in a 2-person scenario, at least a 4-meter localization device is needed when a user requires the labeling accuracy higher than 90\%, but in a 5-person scenario, a 2-meter resolution is required to achieve the same-level accuracy.


\subsubsection{Varying Localization Resolution}
Different techniques have varying performance in localizing individuals, and localizing multiple residents simultaneously is challenging. 
In this experiment, we evaluate how the varying localization resolutions affect the final labeling accuracy in multi-occupancy scenarios.
In each scenario, we implement the same activity sequence in the proposed floorplan but increase the localization resolution. 
We increase the deviations of the residents' trajectories and add more noise to the \textit{best route}.
As shown in Figure~\ref{fig:eloc}, different declining sigmoid curves indicate how the labeling accuracy decrease with the increasing resolution. 

\subsubsection{Effect of Resident Quantity}

In order to understand the decreasing trend of labeling accuracy in different occupancy scenarios, we increase the number of residents in the same setting and investigate the change in labeling accuracy.
Figure~\ref{fig:overall} shows the overview of the labeling performance when increasing the number of residents; Figure~\ref{fig:2&3} and Figure~\ref{fig:4&5} demonstrate the decline rate of the labeling accuracy for the four multi-occupancy scenarios, respectively.
Labeling accuracy decreases when the number of residents increases.
It is worth to note that the highest point of decline rate in these scenarios are all between 3 and 3.5 meters.
The similarity between these four scenarios is because we utilize a similar sensor density for them to keep the same complexity of the sensor deployments.
In other words, the highest point of decline rate depends on the sensor deployments.

\subsubsection{Effect of Sensor Density}
In our experiments, the four multi-occupancy scenarios are using four different floorplans as we described in Table~\ref{tab:multi}, the number of \textit{bedrooms} is changing with the residents' quantity. 
However, we keep the same sensor density of these floorplans, as also shown in Table~\ref{tab:multi}, the sensor density of four scenarios are ranging from 0.43 to 0.46 sensors per m$^2$.
We also utilize \textit{mean distance} to compare the sensor density of four floorplans.
The \emph{Delaunay triangle method}~\cite{an2020Unravelling} is adopted to connect each sensor node with neighboring sensor nodes, while the length of theses connections is leveraged to calculate \textit{mean distance}. 
The distributions of the connections' length are shown in Figure~\ref{fig:sensordistribution}.
After removing outlier-connection near the border, we calculate the average length for these connections.
This average length refers to \textit{mean distance}.
In this setting, same decreasing trends in all scenarios (shown in ~\ref{fig:2&3} and ~\ref{fig:4&5}) indicate the effect of the same sensor density (0.43-0.46 sensor/m$^2$) of our floorplans.
For instance, each sensor occupied 2.3m$^2$ in the 2-person scenario on average, and the mean distance of nodes is 1.6865 meters, where the labeling accuracy decrease most dramatically changed during 3.0-3.1 meters. 
Since other scenarios also have similar sensor density (shown as \textit{mean distance} in Figure~\ref{fig:sensordistribution}), the intervals of transition point for these scenarios are also similar.
This impact is significant when a smart home designer considers the potential sensor layout to better fit the user's requirements.

\section{\emph{M\lowercase{o}S\lowercase{en's}} Sensor selection Strategy}
\label{sec:strategy}

\begin{figure*}[t]
     \centering
     \begin{subfigure}[c]{0.95\textwidth}
         \centering
         \includegraphics[width=\textwidth]{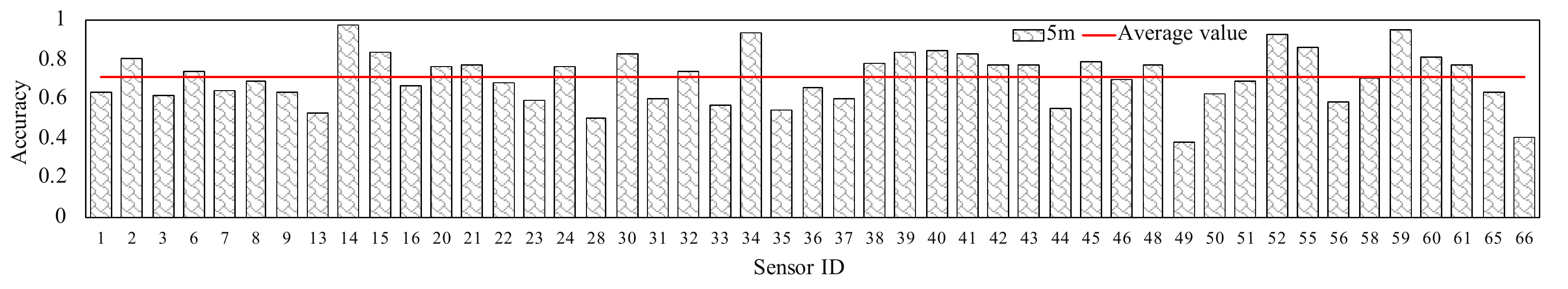}
         \caption{Example of the way each sensor affects the final annotation accuracy; in a 5-person scenario with 5-meter resolution. The red line represents the overall accuracy, comparing to different sensors' performance (grey bars), individually.}
         \label{fig:senone}
     \end{subfigure}
     \begin{subfigure}[c]{0.95\textwidth}
         \centering
         \includegraphics[width=\textwidth]{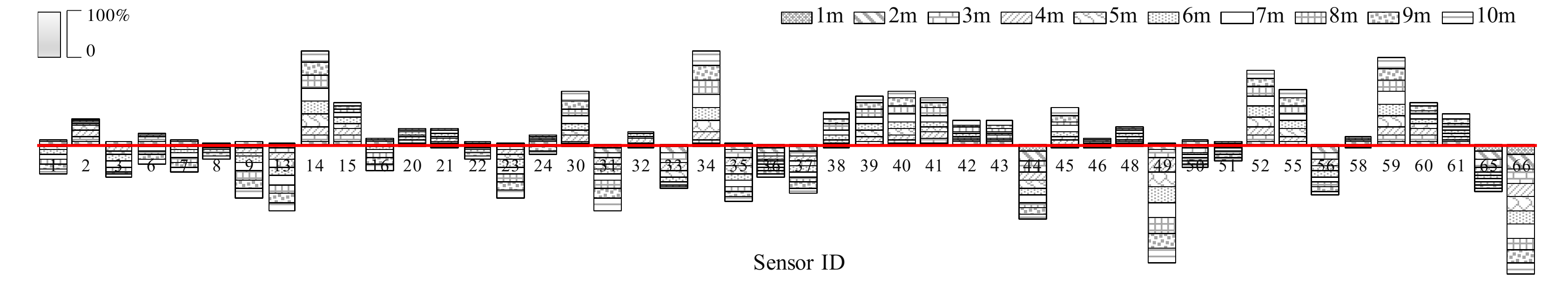}
         \caption{The effect of all sensors in a 5-person scenario, from 1-meter to 10-meter resolution. 
         Each pattern painted block represents the numerical difference compared to the average performance in the respective scenario. The grey block in the corner represents the scale.
         }
         \label{fig:sencum}
     \end{subfigure}
        \caption{Effects on each sensor to overall accuracy in the five-person scenario. 
        }
        \label{fig:sensordifferent}
\end{figure*}

In the sensor-based activity recognition, different sensing systems are proposed to monitor indoor activities by leveraging various ambient sensors, which also have diverse performance.
Previous works~\cite{min2019closer, keally2011pbn} have emphasized the significance of sensor selection on multi-device environments (MDE).
Dynamically selecting the best sensor for a specific activity recognition system is a typical approach developed for context-aware MDE~\cite{zappi2008activity,lee2013active, keally2011pbn}. 
These works are mostly about body sensor network (BSN), and they proposed dynamic designing strategies in the BSN environment.

In this paper, realizing recommendations on sensor selection for different smart homes is the key objective of our analysis, to achieve a better trade-off between user's requirement, labeling accuracy and system cost.
Four requirements (namely cost, acceptance, accuracy and privacy~\cite{chernbumroong2013elderly}) are considered in this work for a practical activity recognition system.
\\
1. \textit{Cost. Low-cost and low-battery consumption of sensors contributes to better competence.}
\\
2. \textit{Acceptance. Wearable or ambient sensors that are non-obtrusive will be better.}
\\
3. \textit{Accuracy. Accuracy is one of the most important factors to be considered.}
\\
4. \textit{Privacy. The non-visual system is preferred.}

The sensor-based activity recognition system is chosen by many researchers due to its non-obtrusiveness and privacy-protection~\cite{wang2019deep}.
The trade-off between the remaining two factors leaves an interesting but tricky balance to attain.
The sensor configuration often depends on the installation cost and sensor prices.
We define \textit{sensor sensitivity} to identify the way a sensor's location and interaction frequency with residents affect the labeling accuracy.
This information is valuable for choosing the best and the most cost-effective sensor network for the smart home environment.
In this section, we focus on identifying \textit{sensor sensitivity} and recommendations on the final sensor selection for a specific layout. 
We use a five-person scenario here as the case study to illustrate the proposed sensor selection strategy. 
The insights from the case study are extendable to other scenarios and \emph{any} new floorplan and sensor layout.

\begin{figure}[t]
     \centering
     \begin{subfigure}[c]{0.23\textwidth}
         \centering
         \includegraphics[width=\textwidth]{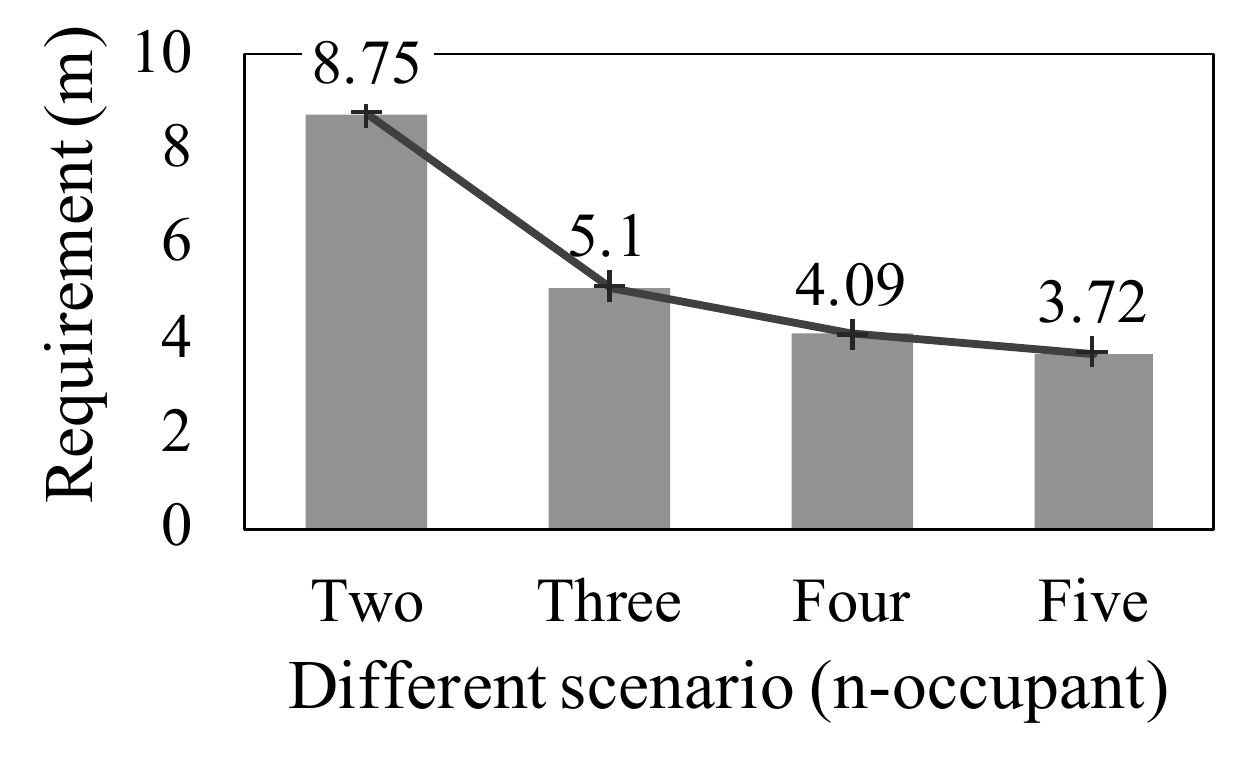}
         \caption{Labeling accuracy at 80\%}
         \label{fig:0.8}
     \end{subfigure}
     \begin{subfigure}[c]{0.23\textwidth}
         \centering
         \includegraphics[width=\textwidth]{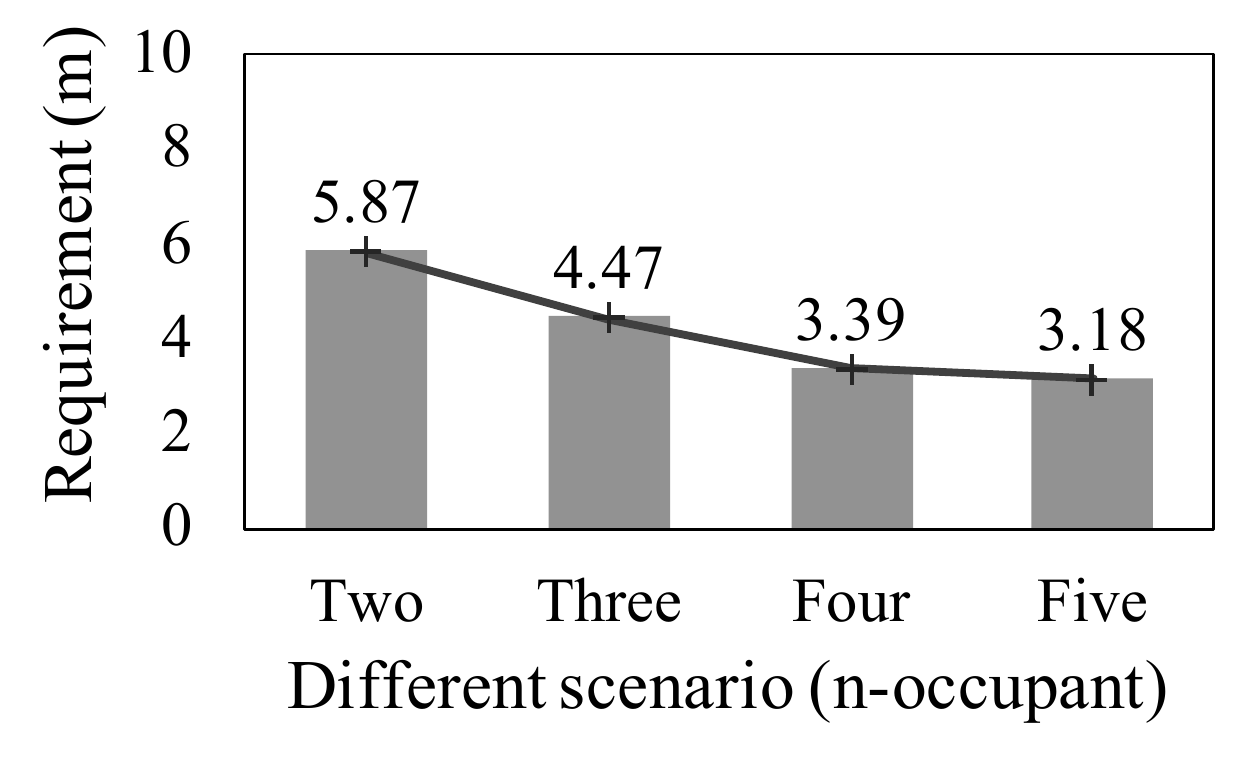}
         \caption{Labeling accuracy at 85\%}
         \label{fig:0.85}
     \end{subfigure}
     \begin{subfigure}[c]{0.23\textwidth}
         \centering
         \includegraphics[width=\textwidth]{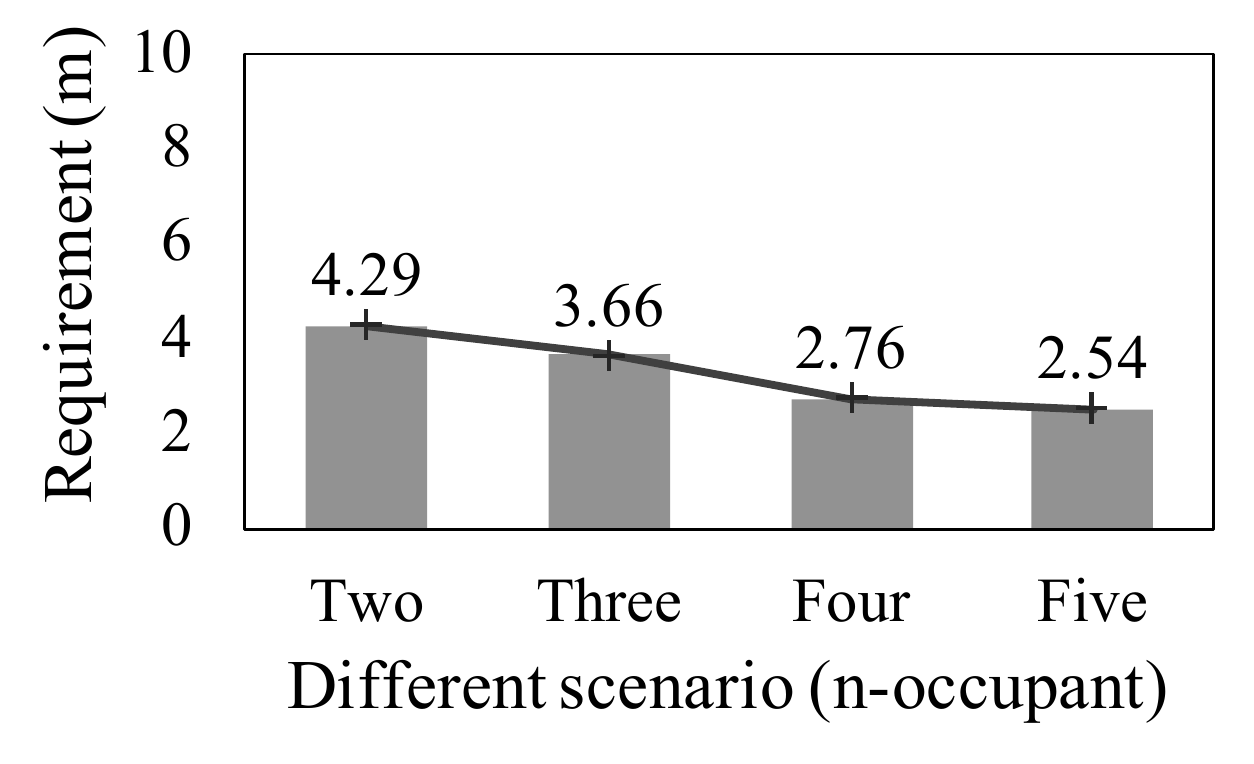}
         \caption{Labeling accuracy at 90\%}
         \label{fig:0.9}
     \end{subfigure}
     \begin{subfigure}[c]{0.23\textwidth}
         \centering
         \includegraphics[width=\textwidth]{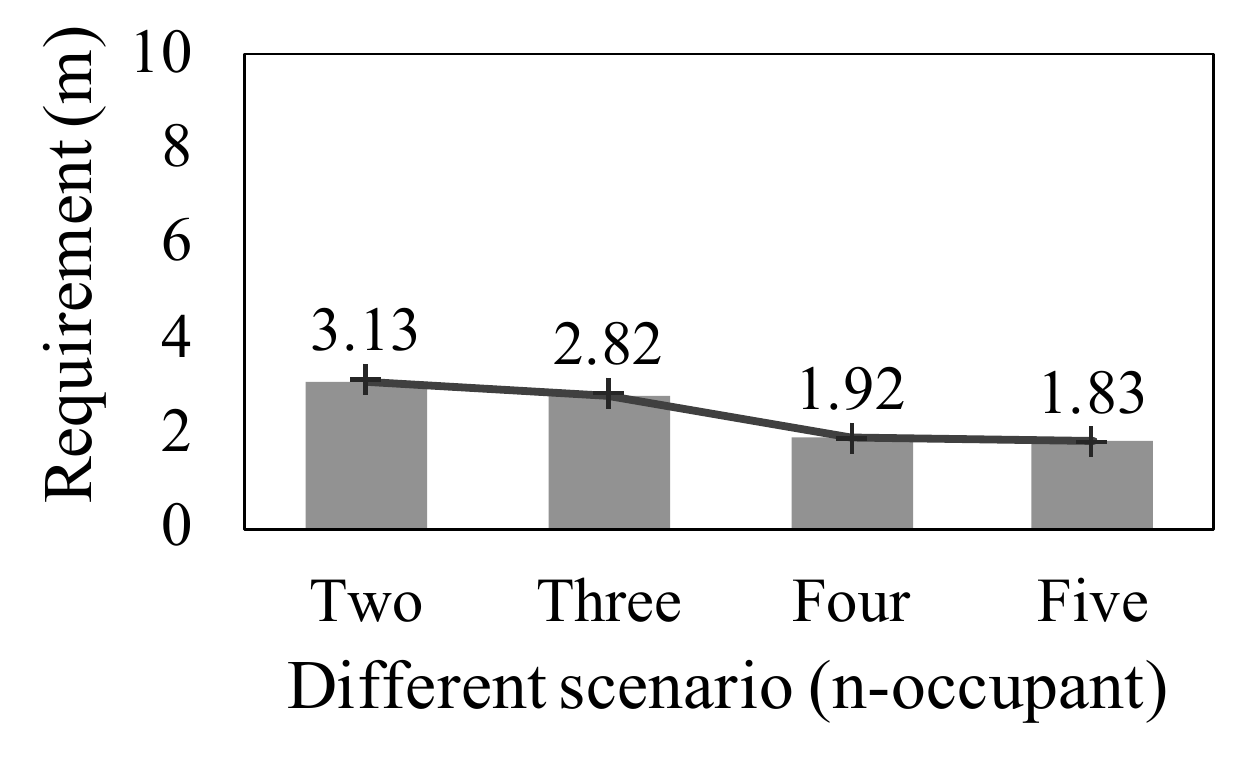}
         \caption{Labeling accuracy at 95\%}
         \label{fig:0.95}
     \end{subfigure}
        \caption{Requirements for localization resolution with the expected labeling accuracy, when the labeling accuracy is 80\%, 85\%, 90\%, 95\%, respectively.}
        \label{fig:requirement}
\end{figure}

\textbf{Identification annotation accuracy.}
The initial analysis of the specific layout (a five-person scenario in this case study) is emulated in the \mosen~system, where annotation accuracy is considered as one of the most important factors in the system.
The related analysis has been described in Section~\ref{sec:evluation}, and the results show how the localization resolution and sensor density affect the accuracy.
Figure~\ref{fig:requirement} shows the requirement for localization resolution with varying labeling accuracy, ranging from 80\% to 95\%, respectively.
For the five-person scenario, if the labeling accuracy is 80\%, the localization resolution requirement should be at least 3.72m or more precise. When the accuracy changed to 90\%, the resolution should attain at least 1.83m, which has higher accuracy but also more expensive than the former one.

\textbf{Sensor individual effect.} 
Each sensor is evaluated in order to identify the individual effect on the integrated performance, as shown in Figure~\ref{fig:sensordifferent}, and the x-axis represents the sensor ID.
With different localization resolutions, sensors might have different performance, and the cumulative effect is shown in this figure.
Each rectangle represents the sensor individual effect under different localization resolution.
For example, Sensor 6 might have opposite effects when the resolution is varying, but for sensor 14, all effects are positive.
This cumulative result leads to a more intuitive concept for \textit{sensor sensitivity}.
Some sensor IDs are absent here as they were not triggered in the experiment.

\begin{figure}[t!]
  \centering
  \includegraphics[width=3.2in]{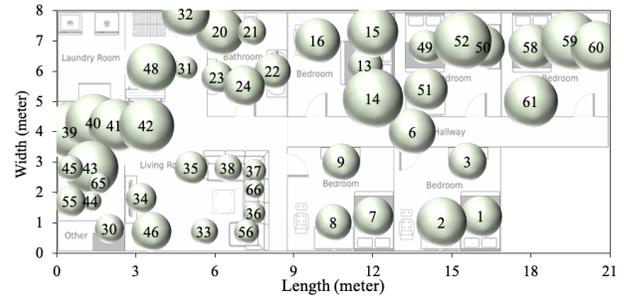}
  \caption{Sensor sensitivity in a five-person scenario. Larger radius of the circle means less sensitive to the distance and allows to have a bigger detection area, vice versa.
  }
  \label{fig:senball}
  \Description{senball.}
\end{figure}

\textbf{Sensor sensitivity.}
A sensor's effect is represented as \textit{sensor sensitivity}, and recommendations will be provided based on sensor sensitivity and sensor cost, as well as the localization resolution and labeling accuracy.
\textit{Sensor sensitivity} integrates the activity patterns of the resident and how frequently residents interact with a specific sensor. 
As shown in Figure~\ref{fig:senball}, the different radii of the circles represent different sensor sensitivities, where a larger radius allows a larger detection area (with lower sensitivity).

\section{discussions}
\label{sec:discussion}
\textit{Data scarcity on multi-occupancy scenarios}
\\
Designing activity recognition systems for multi-occupancy scenarios has been challenging for researchers. 
First, the complexity of human activity increases dramatically when there is more than one person within the same environment. 
Different from the single-occupancy scene, the interactions between residents introduces uncertainty when defining indoor activities.
Second, annotating the triggered sensor with corresponding identification and activity is challenging in the multi-resident scenario. 
The lack of ground-truth values is deficient to the multiple-analysis with advanced machine learning or deep learning technologies.
Third, data privacy is a big concern when collecting real human data. 
Even for the sensor-based activity recognition system, which does not invade privacy as severely as the video-based systems do, there still is a need to attain the trade-off between data utility and privacy.
Fourth, there are still numerous challenges that need to be overcome in the single-occupancy environment~\cite{benmansour2015multioccupant}.
These gaps, hence, impede the practical data collection on the multi-occupancy scenario. 
\\
\textit{Generality of \mosen~System}
\\
We have built \mosen~system to investigate the multi-occupancy scenarios by generating synthetic multi-occupancy behavior model based on real human activity patterns and emulate these models in a virtual environment. 
We use collected datasets from real installations to represents activity patterns of individuals. 
However, available datasets often do not have the direct interaction between residents. 
Due to the data scarcity of the multi-occupancy scene, the synthetic method bridges the gap and the analysis is valuable for us to design a real multi-occupancy scenario in the future.
The strategies proposed in \mosen~system can extend to \emph{any} floorplan, and initial analysis of each specific scenario provide designers on how to better design a sensor-based system in balancing the cost and accuracy.
We choose the identification labeling as the key question in this paper and leave other parameters for future research.
\\
\textit{Towards Practical Utility of Sensor-based system}
\\
Human activities are hard to model in a uniform way, especially when they have different backgrounds, diverse habits, and varied activity performances~\cite{zhan2019activity}.
The uncertainty from the spatial and temporal difference also increases this difficulty~\cite{cook2012casas}.
In the multi-occupancy scenario, activity recognition becomes more sophisticated and challenging with the increasing number of residents.
There is a also a trade-off between the RTLS localization resolution and sensor costs.
Often, researchers in a lab setting prefer the best technology with the highest accuracy, while the accumulated cost are hard to afford in real home designs.
Finally the floorplans and furniture are diverse between different homes, which results in highly diverse sensor layouts in a real environment. 
Our proposed system enables reasonable evaluations and design recommendations for each particular home.

\section{Conclusions}
\label{sec:conclusion}

Often, researchers in a lab setting prefer the best technology with the highest accuracy, while the accumulated cost is hard to afford in real home designs.
The floorplans and furniture are diverse between different designs, which results in highly diverse sensor layouts in the real environment. 
In this paper, we presented \mosen, a framework towards accelerating the real implementation of sensor-based activity recognition systems by analyzing the trade-off between the overall system performance and cost.
We investigated the multi-occupancy scenarios by emulating the synthetic multi-occupancy behavior models, which are generated by real single human activity patterns, in a virtual environment. 
The \mosen~platform can extend to \emph{any} floorplan or sensor configuration. 
The initial analysis of different specific sensor configurations will provide the designers or practitioners with an effective sensor selection strategy.
More quantified results will be shown in future work.

In this paper, we evaluate the efficacy of the \mosen~platform with an automatic identification annotation task using experiments on synthetic datasets and show how the annotation accuracy is affected by the number of residents, different localization resolutions, and sensor density.
Through our trace-driven simulations, the effect of each sensor is also analyzed. Then the sensor selection strategy on the system is provided.
Other context-aware tasks will be emulated in our future work.

\bibliographystyle{ACM-Reference-Format}
\bibliography{MOSen}

\end{document}